\documentclass[showpacs,amsmath,nofootinbib,amssymb,twocolumn,superscriptaddress]{revtex4-1}
\usepackage{array}
\usepackage{color}
\usepackage{dcolumn}
\usepackage{bm}
\usepackage{epsf}
\usepackage{graphicx}
\usepackage{amsmath} 
\usepackage{dsfont} 
\usepackage{hyperref}

\newcommand{\beq}{\begin{equation}}
\newcommand{\eeq}{\end{equation}}
\newcommand{\bea}{\begin{eqnarray}}
\newcommand{\eea}{\end{eqnarray}}

\newcommand\eqn[1]{(\ref{#1})}      % parentheses around the LaTex "ref" macro
\newcommand\Eqn[1]{Eq.~(\ref{#1})}  % includes ``Eq.'' in front
\newcommand\lp{\left(}
\newcommand\rp{\right)}

\newcommand\ctq{\cos{\theta_q}}
\newcommand\stq{\sin{\theta_q}}
\newcommand\cpq{\cos{\phi_q}}
\newcommand\spq{\sin{\phi_q}}
\newcommand\ctp{\cos{\theta_p}}
\newcommand\stp{\sin{\theta_p}}
\newcommand\cpp{\cos{\phi_p}}
\newcommand\spp{\sin{\phi_p}}
\newcommand\nub{\bar{\nu}}

\newcommand\elq{\ell_q}

\begin{document}
\topmargin-0.5in
\textheight 8.7in 
\bibliographystyle{apsrev}
\title{Helicity coherence  in binary neutron star mergers and non-linear feedback}

\author{Am\'elie Chatelain}
\email{chatelai@apc.univ-paris7.fr}
\affiliation{Astro-Particule et Cosmologie (APC), CNRS UMR 7164, Universit\'e Denis Diderot,\\ 10, rue Alice Domon et L\'eonie Duquet, 75205 Paris Cedex 13, France.}

\author{Cristina Volpe}
\email{volpe@apc.univ-paris7.fr}
\affiliation{Astro-Particule et Cosmologie (APC), CNRS UMR 7164, Universit\'e Denis Diderot,\\ 10, rue Alice Domon et L\'eonie Duquet, 75205 Paris Cedex 13, France.}

\begin{abstract}
Neutrino flavor conversion studies based on astrophysical environments usually implement neutrino mixings, neutrino interactions with matter and
neutrino self-interactions. In anisotropic media, the most general mean-field  treatment includes neutrino mass contributions as well, that
introduce a coupling between neutrinos and antineutrinos termed {\it helicity} or {\it spin coherence}. 
We discuss resonance conditions for helicity coherence for Dirac and Majorana neutrinos. 
We explore the role of these mean-field contributions on flavor evolution in the context of a binary neutron star merger remnant.
We find that resonance conditions can be satisfied in neutron star merger scenarios while adiabaticity is not sufficient for efficient flavor conversion. 
We analyse our numerical findings by discussing general conditions to have multiple MSW-like resonances, in presence of non-linear feedback, in astrophysical environments. 
\end{abstract}

\date{\today}


\maketitle

%%%%%%%%%%%%%%%%%%%%%%%%%%%%%%%%%%%%%%%%%%%%%%%%%%%%%%%%%%%%%%%%%%%%%%%%%%%%%%%%
\section{Introduction}
Since the neutrino oscillation discovery by the Super-Kamiokande and the SNO experiments \cite{Fukuda:1998mi,Ahmad:2002jz},
precision measurements have accurately determined the fundamental oscillation parameters. Upcoming experiments should uncover
the neutrino mass ordering, the octant of the second mixing angle, the CP violating phases. 
The combined analysis of available data has a preference for the normal hierarchy, a non-maximal octant and a CP violating phase around $1.5 \pi$, although the statistical significance is still low \cite{Marrone2016}.
Crucial open questions are also the neutrino absolute
mass value and the neutrino nature (Majorana versus Dirac).  
 
 Neutrino propagation in our Sun is modified by neutrino-matter interactions.
These are well accounted for by a mean-field that produces the 
Mikheev-Smirnov-Wolfenstein suppression of high energy solar electron neutrinos \cite{Wolfenstein:1977ue,Mikheev:1986gs}
(see e.g. \cite{Robertson:2012ib,Maltoni:2015kca}). A key achievement 
of solar neutrino physics is also the confirmation
of hydrogen burning of main sequence stars through the proton-proton reaction chain.
This has been crowned by the observation of pp neutrinos by the Borexino collaboration. 

Flavor evolution in dense astrophysical environments has turned out to be complex.
The presence of sizeable neutrino self-interactions render the question of neutrino evolution
intrinsically non-linear and of many-body nature, as first pointed out by Pantaleone \cite{Pantaleone:1992eq}.
The inclusion of $\nu$-$\nu$ interactions \cite{Duan:2006an} has triggered a decade of theoretical investigations.
Models of increasing complexity have revealed a variety of flavor instabilities, some of which have been interpreted in terms
of collective conversion phenomena  (see e.g. the reviews \cite{Duan:2010bg,Chakraborty:2015tfa}).
Flavor instabilities due to the neutrino self-interaction 
occur in core-collapse supernovae, in accretion-disks around black holes \cite{Malkus:2012ts} and compact binary systems,
including black hole-neutron star \cite{Malkus:2014iqa,Malkus:2015mda} and neutron star-neutron star binaries \cite{Zhu:2016mwa,Frensel:2016fge}. 
Such studies are necessary to assess the actual impact on the supernova dynamics and on the nucleosynthetic $r$-process abundances in
neutrino-driven winds in these astrophysical sites.  Observations of future core-collapse supernova explosions,
or of the diffuse supernova neutrino background require a solid understanding of flavor evolution in media as well. 

Theoretical investigations of neutrino propagation in dense media are based on the Liouville Von-Neumann equations
for one-body neutrino density matrices \cite{Sigl:1992fn,Balantekin:2006tg}. However, the most general mean-field equations
include supplementary contributions coming from the non-zero neutrino mass \cite{Volpe:2013jgr,Vlasenko:2013fja,Serreau:2014cfa} and from neutrino-antineutrino pairing correlations \cite{Volpe:2013jgr,Serreau:2014cfa}. Ref.\cite{Vlasenko:2013fja} has first provided quantum kinetic equations for Majorana neutrinos including 
the coupling between neutrinos and antineutrinos due to the neutrino mass. This is referred to as {\it spin coherence} Ref.\cite{Vlasenko:2013fja}  or {\it helicity coherence} Ref. \cite{Serreau:2014cfa}. Ref.\cite{Volpe:2013jgr} has  obtained a rigorous derivation of the neutrino evolution equations based on the Born-Bogoliubov-Green-Kirkwood-Yvon hierarchy approach. Moreover neutrino-antineutrino
pairing correlations have been first implemented explicitly. The concise quantum field theory derivation of Ref. \cite{Serreau:2014cfa} has provided 
the mass and the pairing mean-field contributions in both the Dirac and the Majorana cases. As in the case of mass contributions, pairing correlations also introduce a coupling between neutrinos and antineutrinos. In both cases anisotropy of the medium is necessary for mass and the pairing contributions to the neutrino Hamiltonian to be non-zero. The role of the mean-field contributions to the neutrino evolution equations given by the mass and the pairing terms still need to be fully assessed. Beyond the mean-field approximation, a third correction to the neutrino evolution equations comes from collisions.
In fact, at the neutrinospheres, a small fraction of the neutrino flux can still propagate along non-forward directions due to last collisions.
A small contribution of a backward flux  can produce significant flavor change, as Ref.\cite{Cherry:2012zw} has shown in the context of a core-collapse supernova schematic model.
Demanding simulations that implement self-consistently collisions, neutrino mixings and mean-field terms in a full Boltzmann treatment are still
missing.

The one-flavor schematic model of Ref.\cite{Vlasenko:2014bva} has shown that {\it spin coherence}  can produce a MSW-like phenomenon between neutrinos and antineutrinos. Under specific conditions, a cancellation between the matter and neutrino self-interaction can fulfill the resonance
condition between the two sectors. 
Moreover the non-linearity of the equations can introduce a non-linear feedback. This has a twofold effect :  
the region where the cancellation (and the resonance) occur can be extended and the adiabaticity of the evolution at the resonance can be increased.
As a result significant swapping of the neutrino and antineutrino fluxes is produced for some choices of the parameters. 
This non-linear mechanism is particularly intriguing since the mass contributions turn out to be suppressed by the ratio of the neutrino mass over energy, as one would naively expect \cite{Vlasenko:2013fja,Serreau:2014cfa}. 
A rough estimate of the size of mass and pairing mean-field terms is given in Ref.\cite{Kartavtsev:2015eva} which has also pointed out that a 
MSW-like phenomenon might be produced by the mass terms, while the MSW-like resonance condition cannot be fulfilled for pairing 
contributions. For the latter, non-linearity could still amplify their effects.

The goal of the present manuscript is to investigate the possible role of {\it helicity coherence} within a two-flavor framework, 
based on detailed astrophysical simulations. The aim is to identify under which conditions the helicity coherence resonance can be
fulfilled and non-linear feedback operate.
We choose to set ourselves in a binary neutron star merger environment and employ the results for the matter density profiles,
the electron fraction and the neutrino luminosities 
from the binary neutron star (BNS) merger  simulation of Ref.\cite{Perego:2014fma}. 
We perform numerical calculations to determine neutrino flavor evolution, oscillation probabilities and the associated adiabaticity parameters through the helicity coherence resonances. We consider three model cases that are representative of the ensemble of the astrophysical conditions we have explored.
In order to interpret our numerical findings,  we provide a simple first-order perturbative analysis of the conditions to have multiple resonances 
induced by a non-linear feedback,  producing efficient flavor conversions.
 We take the cases of the matter neutrino resonance (MNR), found in accretion disks around black-holes or binary neutron star mergers, and the model of Ref.\cite{Vlasenko:2014bva} as examples of situations where this mechanism operates and discuss, comparatively, the situation with helicity coherence in our setting. 

The paper is structured as follows. Section II introduces the helicity structure of the Hamiltonian and  of the
mean-field evolution equations, with mass contributions, both  for Majorana and Dirac neutrinos.
Then our schematic model is described and the geometrical factors given.
Section III presents the two-flavor results on
the neutrino flavor evolution.
The resonance and the adiabaticity conditions are discussed for three model cases.  Section IV provides a lowest-order linear analysis
of multiple crossings induced by non-linear feedback. Section V is our conclusions.

\section{Theoretical framework} 
\subsection{Mean-field evolution equations with mass contributions}
\noindent
Neutrino evolution in an astrophysical background of matter, neutrinos and antineutrinos
is usually described through two-point correlation functions. In the case of Majorana neutrinos, these are given by
(we will employ natural units $\hbar=c=k_B=1$)
\begin{align}\label{e:rho}
 \rho_{ij}\left(t,\vec q,- \right)&=\langle a^\dagger_j(t,\vec q,-)a_i(t,\vec q,-)\rangle,\\ \nonumber
 \bar\rho_{ij}\left(t,\vec q,+ \right)&=\langle a^\dagger_i(t,\vec q,+)a_j(t,\vec q,+)\rangle,
\end{align}
and depend on time, on particle momentum and on positive ($h=+$) or negative ($h=-$) helicity states.
The labelling $i,\,j$ are either  mass or flavor indices.
The creation and annihilation operators $a^{\dagger}$ ($b^{\dagger}$)  and 
$a$ ($b$) for neutrinos (antineutrinos) satisfy the nonzero equal-time anti-commutation 
relations 
\beq\label{e:commutators1}
\{ a_{i}(t,\vec p,h), a^{\dagger}_{j}(t,\vec p\,',h') \} 
 = (2 \pi)^3 \delta^{(3)}(\vec p - \vec p\,')\delta_{hh'}\delta_{i j},
\eeq 
and similarly for the anti-particle operators.

In the present manuscript, we will refer to mass contributions to denote corrections to the relativistic limit, that are proportional to the mass and 
associated with  two-point correlators 
\beq\label{e:rhopm}
\zeta_{ij}(t,\vec q\,)=\langle a^\dagger_j(t,\vec q,+)a_i(t,\vec q,-)\rangle,
\eeq
and its hermitian conjugate, that account for helicity change \cite{Vlasenko:2013fja,Serreau:2014cfa}.
Obviously, neutrino evolution studies  also include the usual mixing terms that depend on the mass-squared differences.
We will not refer to these when discussing effects from mass contributions, although they are included in our simulations.

Neglecting collisions, the most general mean-field Hamiltonian has the form \cite{Serreau:2014cfa}
\beq\label{e:Heff}
H_{\rm eff}(t) = \int \mathrm d^3 {x}\, \bar{\psi_i}(t,\vec{x})\Gamma_{ij}(t,\vec{x}) \psi_{j}(t,\vec{x}),
\eeq
that is, it is bilinear in the neutrino field $\psi_i$.
The kernel $\Gamma_{ij}$ 
depends on the composition of 
the background and the kind of interactions that neutrinos undergo. 
In our schematic model in two flavors, we will consider a homogeneous, unpolarised, anisotropic
background of neutrons, protons, electrons, neutrinos and antineutrinos.
Therefore the kernel $\Gamma_{ij}$ of Eq.(\ref{e:Heff})
implements charged- or neutral-current interactions of
$\nu_e$, $\bar{\nu}_e$, $\nu_x$ and $\bar{\nu}_x$\footnote{$\nu_{x} (\bar{\nu}_x)$ stands for muon or tau (anti)neutrinos, or their combination.} with
such particles. 

To render the manuscript self-consistent,
we quote here results that are relevant for the investigation of the effects from mass contributions; while the explicit expression of $\Gamma$
and the detailed derivation of neutrino evolution equations are given in Ref.\cite{Serreau:2014cfa}.
We will present results for Majorana neutrinos, while those for Dirac are reported in Appendix A.

Using the Ehrenfest theorem, extended mean-field equations of motion for
correlators (\ref{e:rho}) and (\ref{e:rhopm}) can be easily derived. For example for the neutrino density matrix one has
\begin{align}\label{e:Ehrenrho}
i \dot{\rho}_{ij}\left(t,\vec q,- \right) = \langle  [a^{\dagger}_{j}\left(t,\vec q,- \right)a_{i}\left(t,\vec q,- \right), H_{\rm eff}(t) ] \rangle. 
\end{align}
Then, the ensemble of equations describing neutrino and antineutrino propagation
can be cast in the compact form\footnote{Note that we have dropped the subscript $M$ used in Ref.\cite{Serreau:2014cfa} everywhere in the equations
to simplify notations.
Note also that in this manuscript we do not consider contributions from the pairing correlators.
The full set of equations for Dirac and Majorana neutrinos with such contributions is given in Appendix A of Ref.  \cite{Serreau:2014cfa}.} \cite{Serreau:2014cfa}
\begin{equation}\label{e:matrixform}
i\, \dot{\!{\rho}}_{\cal G} \left(t,\vec q\,\right) = \left[ {h_{\cal G}}\left(t,\vec q\,\right),{\rho_{\cal G}}\left(t,\vec q\,\right)\right],
\end{equation}
where the generalized density matrix is\footnote{Note that in the present manuscript we denote with $\vec{q}$  instead of $-\vec{q}$ the momentum for antineutrinos.
This former convention introduces sign differences in the expressions where antineutrino momenta are present, 
compared to Ref.\cite{Serreau:2014cfa}, where the latter convention was employed.}
\beq\label{e:Rgen}
\rho_{\cal G} \left(t,\vec q\,\right)
 \!\equiv\!
 \left(\begin{tabular}{cc}
 $\rho\left(t,\vec q\,\right)$&$\zeta\left(t,\vec q\,\right)$\\
 $\zeta^\dagger\left(t,\vec q\,\right)$&$\bar \rho^T\left(t,\vec q\,\right)$
\end{tabular} \right),
\eeq
with $\rho$ and $\bar{\rho}$ $N_f\times N_f$ sub-matrices, corresponding to the usual
neutrino  and antineutrino  density matrices Eq.(\ref{e:rho}). $N_f$ is the number of neutrino families   and
the upper-script $¬T$ indicating transposed.
The generalized Hamiltonian is
\beq
\label{eq:HacheMajo}
h_{\cal G}\left(t,\vec q\,\right)
 \!\equiv\!
   \left(\begin{tabular}{cc}
 $H(t,\vec q\,)$&$\Phi(t,\vec q\,)$\\
 $\Phi^\dagger(t,\vec q\,)$&$-\bar H^T(t,\vec q\,)$
\end{tabular} \right).
\eeq
Both matrices have a $2N_f\times 2N_f$ flavor (or mass) and helicity structure.
The quantities $H$ and $ \bar H$ are the neutrino and antineutrino Hamiltonians respectively, while the off-diagonal term
$\Phi$ is the {\it helicity coherence} matrix, coupling the neutrino and antineutrino sectors.

In the mass basis, the mean-field Hamiltonian contributions are given by
\begin{align} \label{eq:hm}
 H(t,\vec q\,)&=S(t,q)-\hat q\cdot\vec V(t)-\hat q\cdot \vec{V}_m(t),
 \end{align}
 for the neutrino sector and
 \begin{align}\label{eq:hbarm}
 \bar H(t,\vec q\,)&=\bar S(t,q)-\hat q\cdot\vec V(t)-\hat q\cdot \vec{V}_m(t),
\end{align}
for the antineutrino sector. 
The quantity $\hat q = \vec{q}/q$ denotes the unit vector pointing in the neutrino momentum direction ($q$ is the modulus of $\vec{q}$).

The  $N_f\times N_f$ scalar $S(t,q)$ and vector  $\vec V(t)$  matrices receive contributions from
the neutrino mixings, the neutrino-matter charged- and neutral-current interactions as well as the neutral-current
neutrino self-interactions. Their explicit expressions in the flavor basis read \cite{Serreau:2014cfa}
\begin{align}
\label{eq:scalar}
 S(t, q)&=h_0(q)+h_{\rm mat}(t)+ h_{\rm self}(t),\\
\label{eq:scalarbar}
 \bar S(t, q)&=-h_0(q)+h_{\rm mat}(t)+ h_{\rm self}(t),
\end{align}
for neutrinos and antineutrinos respectively.
The first terms correspond to the vacuum  contributions which are 
\beq\label{e:hvac}
h_0 = \mathrm{U} h_{\rm vac} \mathrm{U}^{\dagger}, 
\eeq
with $h_{\rm vac} = diag(E_i)$, $E_{i, i = 1, N_f}$ being the eigenenergies of the propagation eigenstates. The quantity $\mathrm{U}$ is the Maki-Nakagawa-Sakata-Pontecorvo (MNSP) $N_f\times N_f$ unitary matrix relating the mass to the flavor basis \cite{Maki:1962mu}.
The second terms in Eqs.\eqn{eq:scalar}-\eqn{eq:scalarbar} are the scalar neutrino-matter  contribution  to the mean-field 
\begin{align}\label{e:hmat}
h_{\rm mat, \alpha \beta}(t)&=\sqrt{2}G_F \delta_{\alpha \beta} \left[n_e(t) \delta_{\alpha e} -\frac{1}{2}n_n(t)\right],
\end{align}
with the particle number density 
\beq
 n_f(t)=2\int {\mathrm d^3 {p} \over{(2 \pi)^3}}\rho_f(t,\vec p\,),
\eeq
$f = e$ and $n$ standing for electrons and neutrons respectively.
Note that, in Eq.\eqn{e:hmat}, both the charged-current neutrino-electron and the neutral-current  neutrino-neutron contributions need to be included.
In fact,  in our investigation, the neutral current term cannot be discarded from the Hamiltonian $h_{\cal G}$ Eq.\eqn{eq:HacheMajo}, as usually done, since its contribution is not proportional to the identity matrix.

The third terms in Eqs.\eqn{eq:scalar}-\eqn{eq:scalarbar} come from neutral-current neutrino-neutrino interactions
\begin{align}\label{e:hnunus}
h_{\rm self}(t) & = \sqrt{2}G_F\!\!\int {\mathrm d^3 {p} \over{(2 \pi)^3}} \left[\rho (t,{\vec p}) - \bar{\rho}(t,{\vec p})\right] + L,  
\end{align}
with $L$ the conserved lepton number in two flavors
\beq\label{e:L}
L = \sqrt{2}G_F ~ {\rm tr} \Big[\!\!\int {\mathrm d^3 {p} \over{(2 \pi)^3}} \left[\rho (t,{\vec p}) - \bar{\rho}(t,{\vec p})\right] \Big],
\eeq
 ${\rm tr}$ indicating the trace.
Note that, again, the trace terms have to be retained. 
The mean-field matrices Eqs.\eqn{eq:hm}-\eqn{eq:hbarm} involve  the vector term
\begin{equation}
\label{eq:vector}
 \vec V(t)  = \vec V_{\rm mat}(t) + \vec V_{\rm self}(t),
 \end{equation}
that receives contributions from the matter-neutrino current
 \begin{equation}
 \vec V_{\rm mat, \alpha \beta}(t) = \sqrt{2}G_F \delta_{\alpha \beta} \left[\vec J_e(t)\delta_{\alpha e} -\frac{1}{2}\vec J_n(t)\right]
 \end{equation}
 and the neutrino-neutrino one
 \begin{equation}
 \vec V_{\rm self}(t) = \sqrt{2}G_F\!\!\!\int {\mathrm d^3 {p} \over{(2 \pi)^3}}\!\Big\{\hat p\,\left[\rho (t,{\vec p}) - \bar{\rho}(t,{\vec p})\right] \!\Big\} + \vec{k}.
 \end{equation}
The particle velocity densities are
\beq\label{e:currents}
\quad\vec J_f(t)=2\int {\mathrm d^3 {p} \over{(2 \pi)^3}}\vec v_f\rho_f(t,\vec p\,),
\eeq
with $\vec v_f=\vec p/E_p^f$, $E_p^f = \sqrt{p^2 +m_f^2}$, and the quantity $\vec{k}$ is
\beq\label{e:k}
\vec{k} =  \sqrt{2}G_F ~Ê {\rm tr} \!\!\!\int {\mathrm d^3 {p} \over{(2 \pi)^3}}\!\Big\{\hat p\,\left[\rho (t,{\vec p}) - \bar{\rho}(t,{\vec p})\right]\!\Big\},
\eeq
where $\hat p = \vec{p}/p$. In Eqs.\eqn{eq:hm}-\eqn{eq:hbarm} the inclusion of mass contributions gives
a supplementary  diagonal term
\begin{align}
\label{eq:Vm}
 \vec V_m(t) & =-\sqrt{2}G_F\!\!\int {\mathrm d^3 {p} \over{(2 \pi)^3}} \Big\{ e^{-i\phi_p}\hat \epsilon_p\, \Omega(t,\vec{p}) \frac{m}{2p} +  {\rm h.c.} \Big\} \\ \nonumber
 & -\sqrt{2}G_F ~{\rm tr}\int {\mathrm d^3 {p} \over{(2 \pi)^3}}  \Big\{ e^{-i\phi_p}\hat \epsilon_p\, \Omega(t,\vec{p}) \frac{m}{2p} +  {\rm h.c.} \Big\},
\end{align}
with
\begin{align}
\label{eq:Vm2}
\Omega(t,\vec{p}) = \zeta(t,\vec p\,)+\bar\zeta(t,\vec p\,).
\end{align}
Finally the off-diagonal {\it helicity coherence}  matrix  reads \cite{Vlasenko:2013fja,Serreau:2014cfa} 
\begin{align}\label{eq:numix}
 \Phi(t,\vec q\,)&=e^{i\phi_q}\hat \epsilon^*_q\cdot\left[\vec V(t)\frac{m}{2q}+\frac{m}{2q}\,{\vec V}^T\!(t)\right],
\end{align}
where $m$ denotes the mass matrix, and $\phi_q$ is the polar angle of the vector $\vec{q}$ in spherical coordinates. This off-diagonal term mixes neutrino and antineutrino evolution.
The contributions in 
Eqs.\eqn{eq:Vm}-\eqn{eq:numix} come from 
the matter and neutrino currents 
perpendicular to the neutrino direction of motion, since the complex vectors
\beq
\label{eq:lightlike}
\quad \epsilon^\mu(\hat p)=\left(\begin{tabular}{c}$0$\\$\hat\epsilon_p$\end{tabular}\right),
\eeq
and $\hat\epsilon^*_p$ span the plane orthogonal to $\vec p$.\footnote{In terms of an oriented triad of real orthogonal unit vectors $(\hat p,\hat p_\theta, \hat p_\phi)$, for instance the standard unit vectors associated to $\vec p$ in spherical coordinates, one has $\hat\epsilon_p=\hat p_\theta-i\hat p_\phi$.  Note that, $\hat\epsilon_p \cdot\hat\epsilon_p=0$, $\hat\epsilon_p\cdot \hat\epsilon_p^*=2$, $\epsilon^\mu(\hat p)\epsilon_\mu(\hat p)=0$, $ \epsilon^\mu(\hat p)\epsilon^*_\mu(\hat p)=-2$ and
that $\epsilon_\mu(-\hat p)=\epsilon_\mu^*(\hat p)$.} 
As expected the mass terms are suppressed by $m/q$. 
Note that,  in the ultrarelativistic limit, the different helicity sectors are decoupled and one recovers the commonly used theoretical description of neutrino propagation in media.

\subsection{The Majorana case with $N_f =2$}
\noindent
We present here our model to explore effects from the mass contributions
on the neutrino propagation in an astrophysical environment. 
We consider Majorana neutrinos within a two-flavor theoretical framework. 
As we will discuss, such results are also representative of the Dirac case.
The neutrino evolution can be determined using Eq.(\ref{e:matrixform}). 
Unless otherwise specified, from now on, all the expressions will be in the flavor basis. 
The  $4 \times 4$ generalized density matrix Eq.\eqn{e:Rgen}  is given by
\beq\label{e:RM}
\rho_{\cal G} \left(t, \vec{q}\, \right)
=  
\left(\begin{tabular}{c|c}
$\rho$ & $\zeta$ \\  [.1cm]
\hline
\rule{0pt}{2.6ex}$\zeta^{\dagger}$ & $\bar{\rho}^T$ \\ 
\end{tabular} \right) =
\left(\begin{tabular}{cc|cc}
$\rho_{ee}$  &   $\rho_{ex}$  & $\rho^{-+}_{ee}$  & $\rho^{-+}_{ex}$ \\
$\rho_{ex}^*$ & $\rho_{xx}$  &   $\rho^{-+}_{x e}$  &  $\rho^{-+}_{xx}$ \\ [.1cm] \hline
 $\rho^{+-}_{ee}$  & $\rho^{+-}_{x e}$ &  $\bar{\rho}_{ee}$ & $\bar{\rho}_{ex}^*$ \\
$\rho^{+-}_{ex}$  &  $\rho^{+-}_{xx}$  & $\bar{\rho}_{ex}$ & $\bar{\rho}_{xx}$ \\
\end{tabular} \right).
\eeq
Note that, to simplify notations, the explicit dependence on the variables $(t,\vec{q}\,)$ is not shown
on the {\it r.h.s.} of the equation.

 For Majorana neutrinos in two flavors, the MNSP matrix reduces to
\beq\label{e:MajU}
\mathrm{U}= \mathrm{V D} = 
\left(\begin{tabular}{cc}
$\cos \theta$ & $e^{i \alpha/2} \sin \theta$ \\
$-\sin \theta$ & $e^{i \alpha/2} \cos \theta $\\
\end{tabular} \right),
\eeq
where $\mathrm{V}$ is a rotation matrix, while $\mathrm{D} =\mathrm{diag}(1,e^{i \alpha}/2)$ with $\alpha$ the (unknown) Majorana phase. The vacuum Hamiltonian in the flavor basis Eq.\eqn{e:hvac} reduces
to the usual form
\beq\label{e:h02f}
h_0 = \omega
\left(\begin{tabular}{cc}
$- c_{2\theta} $ & $ s_{2\theta} $\\
$s_{2\theta}$ & $ c_{2\theta} $ \\
\end{tabular} \right),
\eeq
with $\omega = {\Delta m^2 \over{4E}} $, $\Delta m^2 = m^2_2 - m^2_1$,  $E = q$ is the neutrino energy, $s_{2\theta} = \sin 2 \theta$ and $c_{2\theta} = \cos 2 \theta$.
The matter term Eq.(\ref{e:hmat}) is 
\begin{align}\label{e:hmat2f}
h_{\rm mat, \alpha \beta}(t)&=  \frac{\sqrt{2}}{2} G_F  \delta_{\alpha \beta} \left[ 2 \delta_{\alpha e} Y_e (t) -  
\left(1-Y_e(t) \right) \right] n_B(t),
 \end{align}
with $n_B$ the baryon number density,
 $Y_e$ the electron fraction.
 Note that we did not include the contributions to the diagonal matrix elements from the matter currents, since they are much smaller than the scalar term.
The neutrino self-interaction Hamiltonian Eqs.\eqn{e:hnunus}-\eqn{eq:vector} reads
\begin{align}\label{e:hnunu2f}
h_{\nu \nu} \left( t, \vec{q}\,\right) & = \sqrt{2}G_F \sum_{\alpha=e,x}  [  \int  (1 - \hat{q} \cdot \hat{p}) \\ \nonumber
&  \times [\mathrm dn_{\nu_{\underline{\alpha}}} \rho_{\nu_{\underline{\alpha}}} (t,{\vec p}) - \mathrm dn_{\bar{\nu}_{\underline{\alpha}}}  \bar{\rho}_{ \bar{\nu}_{\underline{\alpha}}}(t,{\vec p})] ]  + L - \hat{q} \cdot \vec{k} , 
 \end{align}
 with $L$ and $\vec{k}$ given by Eqs.\eqn{e:L} and \eqn{e:k} respectively.
The quantity $dn_{\nu_{\underline{\alpha}}}$ denotes the differential number density of neutrinos and the underline refers to the neutrinos initially born with $\alpha $ flavor.
Besides such contributions that are usually included in flavor evolution studies, the Hamiltonian presents the diagonal mass term 
and the off-diagonal one that depends on the matter and the neutrino currents. 
As we will discuss, since the diagonal contribution from the neutrino mass Eq.(\ref{eq:Vm})  is very small, it will not be implemented in our calculations.

The generalized Hamiltonian matrix  Eq.\eqn{eq:HacheMajo} reads\footnote{Here we have omitted again the explicit dependence on the variables not to overburden notations.}
\begin{widetext}
\beq
\label{eq:HM}
h_{\cal G} \left(t, \vec{q}\, \right)
=  \left(\begin{tabular}{cc|cc}
$- \omega c_{2\theta} + \lambda Y_e  + h^{ee}_{\nu \nu}  $ &  $ \omega s_{2\theta} + h^{ex}_{\nu \nu}  $ & 
 $\Phi_{ee}$ &   $\Phi_{ex}$ \\
$ \omega  s_{2\theta} + h^{xe}_{\nu \nu}$ &   $\omega c_{2\theta} + h^{xx}_{\nu \nu}  $ &  $ \Phi_{xe}$&   $\Phi_{xx}$ \\ [.1cm]  \hline
\rule{0pt}{2.6ex} $ \Phi^\dagger_{ee}$ & $\Phi^\dagger_{ex}$ & $- \omega  c_{2\theta} +  \lambda (1 - 2 Y_e)  - h^{ee}_{\nu \nu} $& $\omega  s_{2\theta}- h^{xe}_{\nu \nu}  $ \\
$\Phi^\dagger_{xe}$ & $ \Phi^\dagger_{xx}$ & $\omega s_{2\theta}- h^{ex}_{\nu \nu}  $ & $\omega c_{2\theta} +  \lambda (1- Y_e)  - h^{xx}_{\nu \nu} $ 
\end{tabular} \right),
\eeq
\end{widetext}
where $\lambda =  \sqrt{2} G_F n_B $.
Note that the quantity $\frac{\lambda}{2} (Y_e - 1) \mathbb{I}_{4 \times 4}$, with $\mathbb{I}_{4 \times 4}$ the identity matrix, has been subtracted from the diagonal.

The helicity coherence terms Eq.\eqn{eq:numix} in the flavor basis is given by 
\beq\label{e:phif}
\Phi(t,\vec{q}) =  \left[ e^{i\phi_q}\hat \epsilon^*_q\cdot \vec{V}(t) \right] \mathrm{U} {m \over{2 q}} \mathrm{U}^{T} + U {m \over{2 q}} \mathrm{U}^T  \left[ e^{i\phi_q}\hat \epsilon^*_q\cdot \vec{V}^T\!(t) \right].
\eeq
By using   $c_\theta = \cos{\theta}$ and $s_\theta=\sin{\theta}$, one can rewrite the factor associated with the mass matrix as  
\begin{align}\label{e:Mdec}
\mathrm{U} {m \over{2 q}} \mathrm{U}^T & = m_0 
\left(\begin{tabular}{cc}
$c^2_\theta  + e^{i \alpha}  s^2_\theta$& $s_\theta c_\theta (e^{i \alpha} -1)$ \\
$s_\theta c_\theta(e^{i \alpha} -1)$ & $s^2_\theta  + e^{i \alpha}  c^2_\theta$\\
\end{tabular}\right) \\ \nonumber
& + {\Delta m^2 \over{4 m_0}} 
\left(\begin{tabular}{cc}
$- c^2_\theta  + e^{i \alpha}  s^2_\theta$& $s_\theta c_\theta(e^{i \alpha} + 1)$ \\
$s_\theta c_\theta(e^{i \alpha} +1)$ & $- s^2_\theta  + e^{i \alpha}  c^2_\theta$
\end{tabular}\right),\label{e:mfl}
\end{align}
where we have introduce the quantity $m_0 = (m_1 + m_2)/2$.

\subsection{The Dirac case with $N_f =2$}
\noindent
We present here the explicit expression of the Hamiltonian in the Dirac case. The equations of motion are given in Appendix A. 
The main difference from the Majorana case is that the sub-sectors with the "wrong" helicities, $\rho_{++}$ and $\bar{\rho}_{--}$, involve sterile components. Moreover, in the Dirac case, there are two $4\times4$ generalized Hamiltonians that need to be evolved : one for neutrinos, and one for antineutrinos.
For neutrino, the generalized density matrix \Eqn{eq:dhelicity1} reads 
\beq\label{eq:helicity1}
\rho_{D,{\cal G}}  \left(t, \vec{q}\, \right) 
=  
\left(\begin{tabular}{c|c}
$\rho$ & $\ \zeta$ \\  [.1cm]
\hline
\rule{0pt}{2.6ex}$\zeta^{\dagger}$ & $\ \tilde{\rho}$ \\ 
\end{tabular} \right) =
\left(\begin{tabular}{cc|cc}
$\rho_{ee}$  &   $\rho_{ex}$  & $\rho^{-+}_{ee}$  & $\rho^{-+}_{ex}$ \\
$\rho_{ex}^*$ & $\rho_{xx}$  &   $\rho^{-+}_{xe}$  &  $\rho^{-+}_{xx}$ \\ [.1cm] \hline
 $\rho^{+-}_{ee}$  & $\rho^{+-}_{ex}$ &  $\tilde{\rho}_{ee}$ & $\tilde{\rho}_{ex}$ \\
$\rho^{+-}_{xe}$  &  $\rho^{+-}_{xx}$  & $\tilde{\rho}^*_{ex}$ & $\tilde{\rho}_{xx}$ \\
\end{tabular} \right).
\eeq
The $\lp -- \rp$ sub-sector in the generalized Hamiltonian \eqn{e:H} is very similar to the one in the Majorana case ; however, due to the fact that the sterile component does not interact with matter or neutrinos, the $\lp ++ \rp$ sub-sector includes only the $2\times 2$ vacuum Hamiltonian. The generalized Hamiltonian for neutrinos is therefore 
\begin{widetext}
\beq
\label{eq:HD}
h_{D,{\cal G}}\left(t, \vec{q}\, \right)
=  \left(\begin{tabular}{cc|cc}
$- \omega c_{2\theta} +  \lambda' (3Y_e-1)   + h_{\nu \nu}^{ee}  $ &  $\omega s_{2\theta} + h_{\nu \nu}^{ex}  $ & 
 $\tilde{\Phi}_{ee}$ &   $\tilde{\Phi}_{ex}$ \\
$ \omega s_{2\theta} + h_{\nu \nu}^{xe}$ &   $\omega c_{2\theta} +  \lambda' (Y_e-1) + h_{\nu \nu}^{xx}  $ &   $ \tilde{\Phi}_{xe}$&   $\tilde{\Phi}_{xx}$\ \\
 &  & &  \\
\hline
& & & \\
$ \tilde{\Phi}^\dagger_{ee}$ & $\tilde{\Phi}^\dagger_{ex}$  & $-\omega  c_{2\theta}$& $\omega  s_{2\theta}$ \\
$\tilde{\Phi}^\dagger_{xe}$ & $ \tilde{\Phi}^\dagger_{xx}$ & $\omega s_{2\theta}  $ & $\omega c_{2\theta}  $ 
\end{tabular} \right),
\eeq
\end{widetext}
with $ 2 \lambda'  =  \lambda$.
A similar expression can be written for the generalized Hamiltonian for anti-neutrinos, $\bar{h}_{D,{\cal G}}$ \eqn{e:Hbar}, with, of course, the $\lp -- \rp$ and $\lp ++ \rp$ sectors reversed.

\subsection{Our schematic model based on neutron star mergers simulations}
\noindent
Neutron star mergers produce lots of low energy neutrinos in the accretion disk during the post-merger phase.
In such sites flavor evolution studies show the presence of MNR conversion phenomena 
that require a cancellation between the matter and the neutrino self-interaction contributions. 
As we will show, the corresponding resonant condition 
 is very close to one of the resonant conditions due to the helicity coherence term.
Moreover, MNR also shows a non-linear feedback mechanism
that presents a similarity, in the sense that it is capable of maintaining the resonance over long distances, with the one found in the first (one-flavor) study of mass effects
in core-collapse supernovae \cite{Vlasenko:2014bva}. In order to explore mass effects in a more realistic settings,
we have  built a two-flavor schematic model in an extended mean-field approximation, based on
 simulations of BNS mergers. Our goal is  to identify if and under which conditions, the mass contributions can produce efficient flavor
conversion.

\begin{figure}
\includegraphics[scale=1]{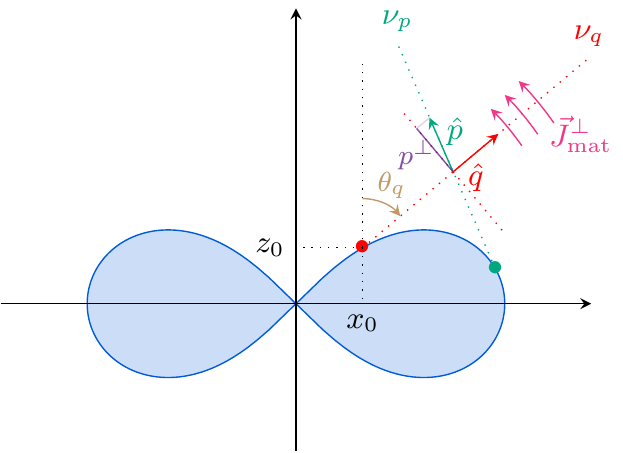}
\caption{Geometry of our model. The blue surface shows the neutrinospheres, that we will approximate later on as infinitely thin disks. We chose the emission point $\lp x_0, z_0 \rp$ of our test neutrino $\nu_q$ on this surface, while the angle $\theta_q$ fixes the direction of its momentum $\hat{q}$ ($\phi_q$ is set to zero). The quantity $\vec{J}^\perp_\text{mat}$ indicates the perpendicular matter current. The momentum $\hat{p}$ of the background neutrino $\nu_p$ also has a component $p^\perp$ perpendicular to the test neutrino, creating a neutrino current perpendicular to the test neutrino trajectory. }\label{fig:schem}
\end{figure}

According to the detailed simulations of Ref.\cite{Perego:2014fma}, 
a central object is formed  by the merging process, with a radius  of about 30 km.
In our scenario neutrinos produced in such event
evolve in a static background of matter, neutrinos and antineutrinos.
Therefore we will replace the $t$-dependence of our variables by an $r$-dependence, i.e. the distance $r$ travelled by the neutrino from its point of emission. 
To simplify the problem while keeping the essential features, we approximate the neutrinospheres as an infinitely thin disk with maximal sizes $R_{\nu}$, as previously done in the literature \cite{Malkus:2012ts,Malkus:2014iqa,Zhu:2016mwa,Frensel:2016fge}.
Three different disk sizes are considered for the $\nu_e$, $\bar{\nu}_e$ and $\nu_{x}$ (or $\bar{\nu}_x$) (Table \ref{tab:fluxes}).
In particular, the $\nu_x$ and $\bar{\nu}_e$ neutrinosphere radii are very close and smaller than the $\nu_e$ outermost radius.
Note that this difference in $R_{\nu}$ and in the luminosities can induce a sign change in the neutrino self-interaction potential,
producing the so-called symmetric matter neutrino resonance (sMNR) phenomenon where both electron neutrinos and antineutrinos modify their flavor content. This phenomenon is first pointed out in in an accretion-disk black hole scenario Ref.\cite{Malkus:2012ts}  and further investigated in \cite{Malkus:2014iqa,Zhu:2016mwa,Frensel:2016fge}. 

Our model is two-dimensional and has an azimuthal symmetry axis (see Figure \ref{fig:schem}). Neutrinos evolve along a straight line trajectory (we neglect the bending due to the presence of strong gravitational fields).
In order to follow neutrino evolution along a given trajectory, we use
a spherical coordinate system given by $(r, \theta, \phi)$ (Figures \ref{fig:schem} and   \ref{fig:def_geometry}), while for the neutrino background
it is useful to express $(\theta,\phi)$ back to the emission point $(r_d, \varphi, 0)$ on the disk, as first introduced in Ref.\cite{Dasgupta:2008cu}à (see Appendix B and Figure \ref{fig:birdeye}). 
For the matter Hamiltonian Eq.\eqn{e:hmat}  we have used 
cylindrical averages of the the electron fraction and the baryon number density results of Ref.\cite{Perego:2014fma}.
Therefore, in our calculations, both $n_B = n_B(r)$ and $Y_e = Y_e(r)$. 
 
\begin{table} 
\center
\begin{tabular}{ccccccc}
~~~~~~~~~&  $\langle E_{\nu} \rangle$ & $L_{\nu}$   & $R_{\nu}$  \hfill \\ \hline
$\nu_{e}$ &  $10.6$ & $15$  & 84 \\
$\bar{\nu}_{e}$ & $15.3$ & $30$ &  60 \\
$\nu_{x}$ & $17.3$ & $8$ &   58 \\
 \hline
\end{tabular}
\caption{Electron and non-electron neutrino flavors : Average neutrino energy ({\rm MeV}) from Ref.\cite{Frensel:2016fge}, luminosities  in units of $10^{51}~{\rm erg/s}$ from Ref.\cite{Perego:2014fma}. 
The last column furnishes the  outermost radii ({\rm km}) Ref. \cite{Frensel:2016fge}. Such values correspond to the neutrinospheres of a neutron star merger at 100 ms after the merging process. }
\label{tab:fluxes}
\end{table}
As for the self-interaction Hamiltonian, one needs to implement the differential number density $\mathrm dn_{\nu_{\underline{\alpha}}}$ 
\beq\label{e:dn}
\mathrm dn_{\nu_{\underline{\alpha}}} =j_{\nu_\alpha}\lp p\rp \mathrm dp \mathrm d\phi_p \mathrm d\cos{\theta_p},
\eeq
for neutrinos emitted isotropically from any point on the surface of the disk. A similar expression holds for antineutrinos.
The quantity 
\beq\label{e:j}
j_{\nu_\alpha} \lp p\rp = \frac{L_{\nu_{\underline{\alpha}}}f_{\nu_{\underline{\alpha}}} \lp p \rp}{\pi^2 R^2_{\nu_\alpha} \left< E_{\nu_\alpha} \right>},
\eeq
is the neutrino number density per unit angle per unit energy, and $\lp \theta_p, \phi_p \rp$ the spherical coordinates of $\hat{p}$ (Figure \ref{fig:def_geometry}). The angular integration is performed over the boundaries $\Omega_{\nu_\alpha}$ ($\Omega_{\bar{\nu}_{\alpha}}$)  of the corresponding $\nu$ ($\nub$) neutrinosphere. 
Introducing Eq.\eqn{e:dn}-\eqn{e:j} into \eqn{e:hnunu2f}  the explicit expression for the neutrino-neutrino term reads
\begin{widetext}
\begin{align}\label{e:hexp}
h_{\nu \nu}(r, q, \ell_q) & =\sqrt{2} G_F   \sum_{\alpha=e,x}  \int_{0}^{\infty}  \int_{\Omega_{\bar{\nu}_{\alpha}, \nu_{\alpha}}} \!\!\!\!
\mathrm dp\ \mathrm d\Omega (1 - \hat{q} \cdot \hat{p})  \left[\rho_{\nu_{\underline{\alpha}}} (r, p, \ell_p)
 {{L_{\nu_{\underline{\alpha}}} f_{\nu_{\underline{\alpha}}}(p) }  \over{ \pi^2 R^2_{\nu_\alpha}\left< E_{\nu_\alpha} \right>}} 
  -   \bar{\rho}_{ \bar{\nu}_{\underline{\alpha}}} (r, p, \ell_p) {{L_{ \bar{\nu}_{\underline{\alpha}}} f_{\bar{\nu}_{\underline{\alpha}}}(p)}  \over{ \pi^2 R^2_{\bar{\nu}_\alpha}\left< E_{\bar{\nu}_\alpha} \right>}} 
  \right], 
\end{align}
\end{widetext}
where the underline in  $\nu_{\underline{\alpha}}$ and $\bar{\nu}_{\underline{\alpha}}$  
indicates the initial neutrino flavor. The variables, on which the neutrino evolution depends, include
 $\ell_i  \!\equiv\!  (\theta_i, \phi_i, \bf{Q_0}) $ with the angles ($\theta_i, \phi_i$) ($i$ = $p$ or $q$) 
defining the neutrino trajectory and the coordinates ${\bf Q_0} \equiv (x_0, z_0)$  giving the neutrino point of emission in the $\pi_{xz}$ plane.
The functions $L_{\nu_{\underline{\alpha}}}$ ($L_{ \bar{\nu}_{\underline{\alpha}}}$) are the total neutrino luminosities, that have to be divided by two in \Eqn{e:hexp} since we consider the neutrino emitted in only one hemisphere, whereas 
$ f_{\nu_{\underline{\alpha}}} $ ($f_{\nu_{\underline{\alpha}}}$) are the neutrino (antineutrino) spectra,  at the neutrinospheres.

In this first exploratory work based on a two-dimensional model for two-neutrino flavors, we have used an approximate treatment of the self-interaction Hamiltonian 
that consists in assuming that neutrino trajectories are all coupled and follow the same flavor history as the test neutrino along a given trajectory, i.e.
\beq\label{e:sa}
\rho_{\nu} (r,\vec{p}) = \rho_{\nu} (r, p), 
\eeq 
and similarly for $\bar{\rho}_{\nu}$.
This procedure is analogous to the so-called "single-angle" approximation in the core-collapse supernova context, first introduced in the 
{\it bulb} model \cite{Duan:2006an}. We emphasise that our treatment of the self-interaction reduces to the "single-angle" approximation,
if one imposes spherical and azimuthal symmetry, as in the {\it bulb} model. 
According to multi-angle studies of flavor evolution in core-collapse supernovae, the inclusion of the full angular dependence of the density
matrices can introduce decoherence of collective flavor conversion effects (see e.g. \cite{EstebanPretel:2007ec}). In case of positive findings in future studies, one would need to go beyond and implement the full angular dependence in  Eq.\eqn{e:sa}.

By imposing Eq.\eqn{e:sa} the integral over the angular variables
can be performed giving the geometrical factor
\begin{align}\label{e:geom}
G_{\nu_\alpha} (r,\ell_q)  = \int_{\Omega_{ \nu_{\alpha}}} \mathrm d \Omega(1 - \hat{q} \cdot \hat{p}) ,
\end{align}
and similarly for $G_{\bar{\nu}_{\alpha}} $. As a consequence, Eq.\eqn{e:hexp} becomes 
\begin{widetext}
\begin{align}\label{e:hexpsa}
h_{\nu \nu} (r,q, \elq)  & =\sqrt{2} G_F  \sum_{\alpha=e,x}
   \int_{0}^{\infty}\! \mathrm dp   \left[ G_{\nu_{\alpha}} (r,\elq)\rho_{\nu_{\underline{\alpha}}}(r,p)  {{L_{\nu_{\underline{\alpha}}} f_{\nu_{\underline{\alpha}}}(p) }  \over{ \pi^2 R^2_{\nu_\alpha}\left< E_{\nu_\alpha} \right>}} 
  -   \bar{\rho}_{ \bar{\nu}_{\underline{\alpha}}}(r,p) 
  G_{\bar{\nu}_{\alpha}} (r,\elq) {{L_{ \bar{\nu}_{\underline{\alpha}}} f_{\bar{\nu}_{\underline{\alpha}}}(p)}  \over{\pi^2 R^2_{\bar{\nu}_\alpha}\left< E_{\bar{\nu}_\alpha} \right>}}  \right]. 
\end{align}
\end{widetext}
The angular variables in \eqn{e:geom} can be expressed as a function of the $(r_d, \varphi)$ variables defining the point in the emission
plane $\pi_{xz}$ (Figure \ref{fig:birdeye}). The integral over $\varphi$ is easily performed and the geometric factor becomes
\begin{align}\label{e:G}
G_{\nu_\alpha}  (r,\ell_q) & = z \int_{0}^{R_{\nu_{\alpha}}} \mathrm{d} r_{\mathrm{d}} \, r_{\mathrm{d}} \, \Gamma(r_{\mathrm{d}}, \elq, r),
\end{align}
where the explicit expression for $\Gamma$ is given by Eqs.(\ref{e:geomg}-\ref{e:Ek}).

For the mass effects, one needs to specify the matter and self-interaction contributions to the 
helicity coherence term \eqn{eq:numix} as well as the supplementary diagonal contribution \eqn{eq:Vm}.
By taking constant matter velocities, 
the matter currents contribution in Eq.\eqn{eq:numix} becomes
\begin{align}
\label{e:betacur}
 \hat{\epsilon}^*_q\cdot \vec V_{\rm mat, \alpha \beta}(r)\! & = \frac{\sqrt{2}}{2} G_F \beta  \delta_{\alpha \beta} \left[ 2 \delta_{\alpha e} Y_e (t) -  
\left(1-Y_e(t) \right) \right] n_B(r).
 \end{align}
For the self-interaction contribution to the helicity coherence term, one needs to calculate $\hat{\epsilon}^*_q\cdot \vec V_{\rm self}(r)$, that is 
\begin{align}
h_{\nu \nu}^\perp(r,q,\ell_q)  &= \sqrt{2} G_F \sum_\alpha \int_0^{\infty} \!\!\!\!\!\! \mathrm dp  \left \{ \int_{\Omega_{\nu_\alpha}} \!\!\!\!\!\! \lp \hat{\epsilon}^* \lp \hat{q} \rp \cdot \hat{p} \rp \rho_{\nu_{\underline{\alpha}}} (r,p,\ell_p) \mathrm dn_{\nu_{\underline{\alpha}}} \right. \nonumber \\
&  -\left. \int_{\Omega_{\bar{\nu}_\alpha}} \!\!\!\!\! \lp \hat{\epsilon}^* \lp \hat{q} \rp \cdot \hat{p} \rp \bar{\rho}_{\bar{\nu}_{\underline{\alpha}}}  (r,p,\ell_p) \mathrm dn _{\bar{\nu}_{\underline{\alpha}}}  \right \}.
\label{eq:hselfperp}
\end{align}

\begin{figure}[!thb] 
    \centering
    \begin{minipage}{.5\textwidth}
        \includegraphics[scale=0.376]{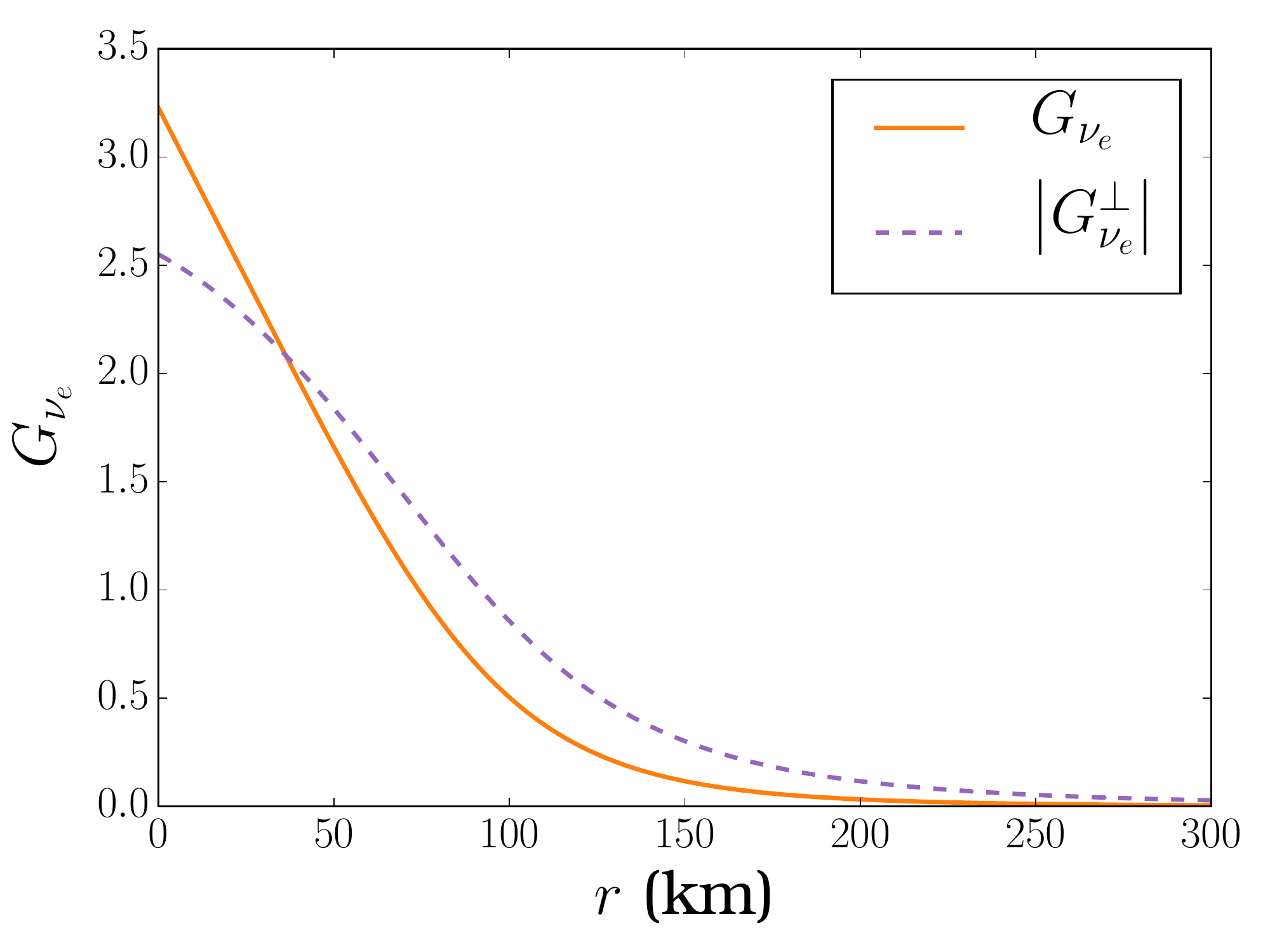}
        \includegraphics[scale=0.376]{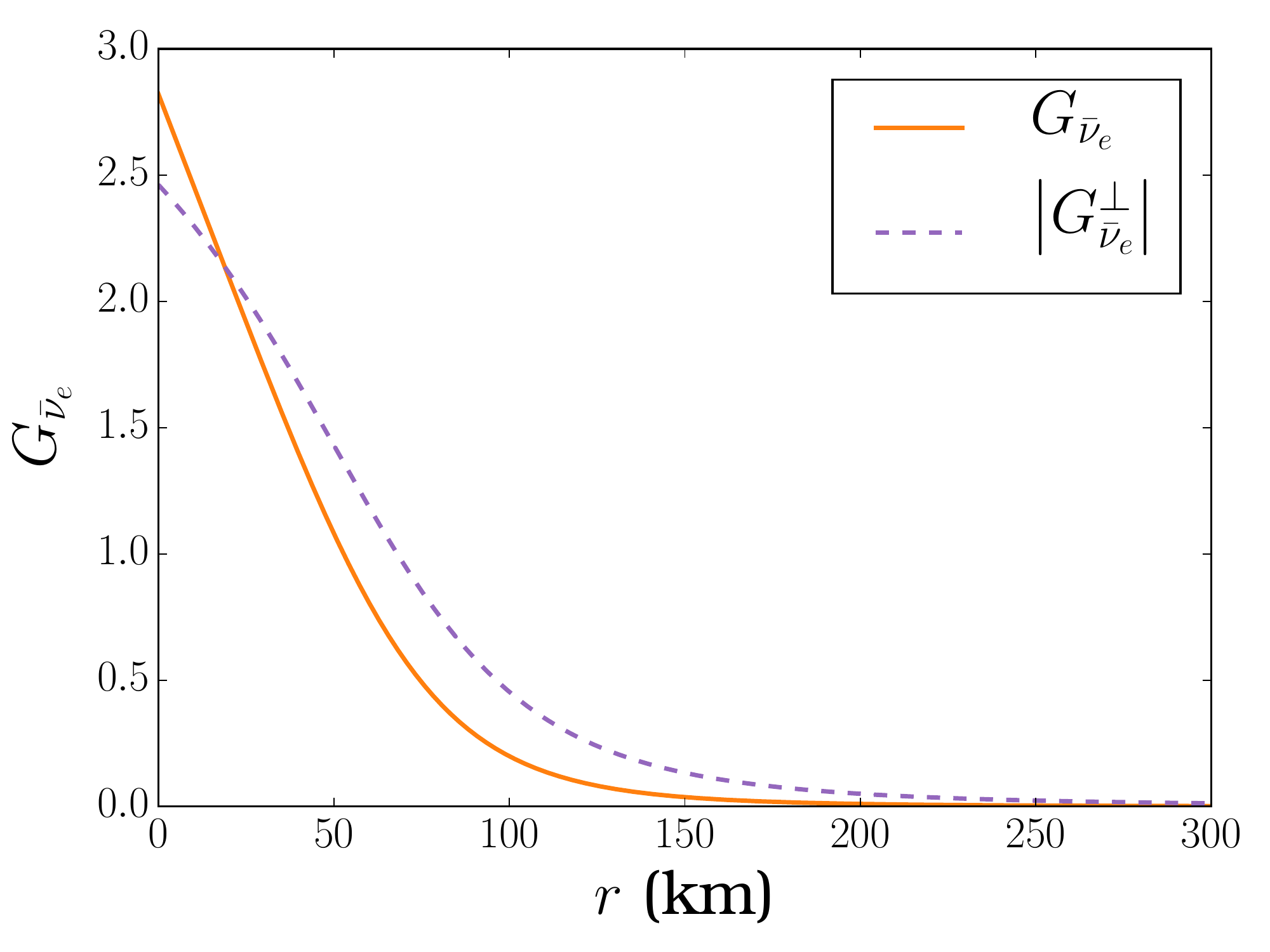}
        \caption{Geometrical factors for neutrinos (upper) and antineutrinos (lower figure),
        as a function of distance, in our schematic model based on binary star mergers. The two curves correspond to  $G_{\nu}$  \eqn{e:G} (solid line)  and $G^{\perp}_{\nu}$  \eqn{e:cperp} (dashed line) in the self-interaction Hamiltonian Eq.\eqn{e:hexpsa} and  Eq.\eqn{eq:hselfperp} respectively. Results correspond to Model C (Table \ref{tab:trajectory}).}
 \label{fig:one}
    \end{minipage}
\end{figure}

\begin{figure}[!thb]
    \centering
    \begin{minipage}{.5\textwidth}
        \centering
        \includegraphics[scale=0.376]{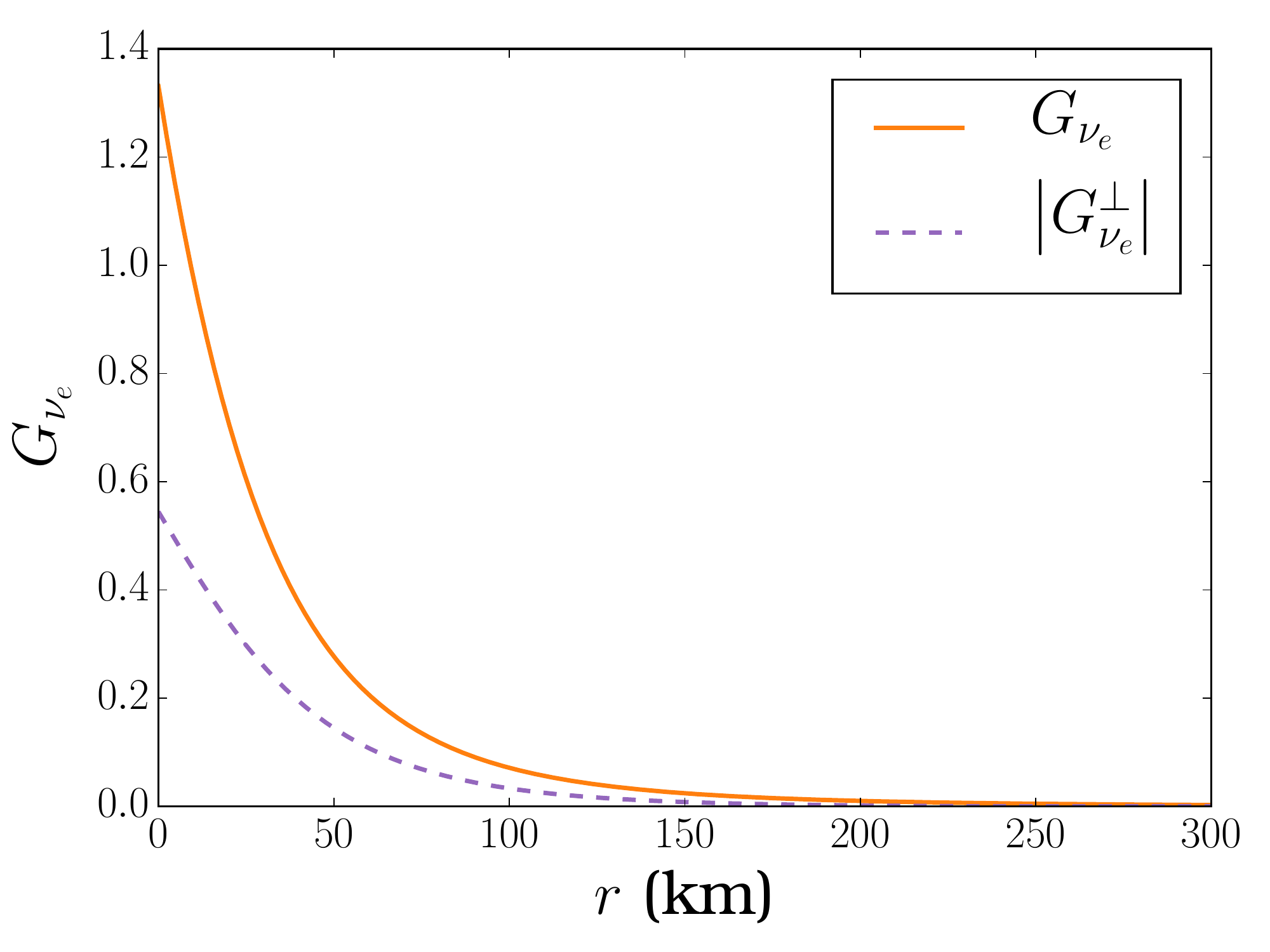}
        \includegraphics[scale=0.376]{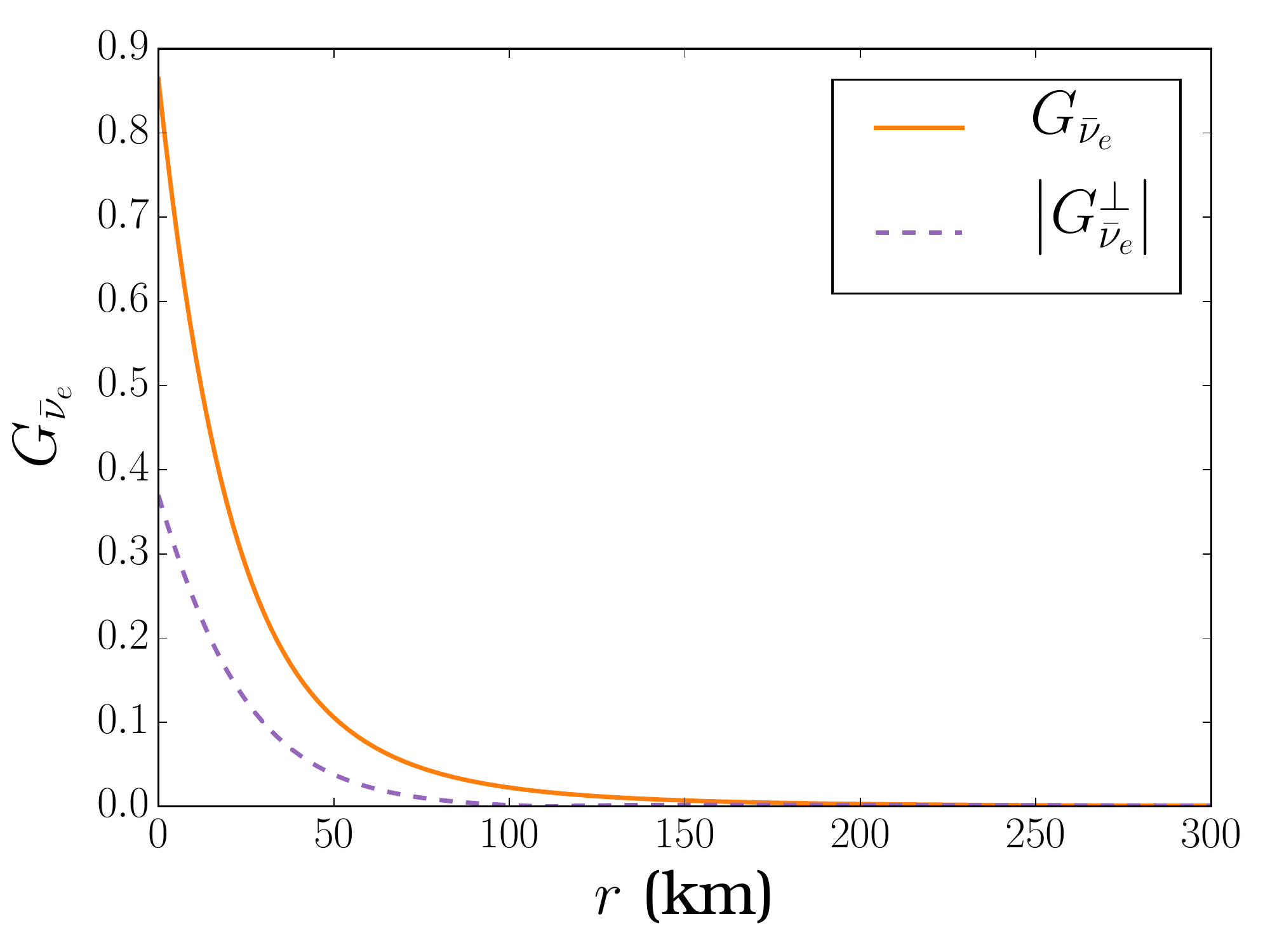}
        \caption{Same as Figure \ref{fig:one} for Model A (see Table \ref{tab:trajectory}).}
\label{fig:two}
    \end{minipage}
\end{figure}

Using the hypothesis \eqn{e:sa}, a perpendicular geometrical factor can be defined as 
\begin{align}\label{e:cperp}
G^\perp_{\nu_\alpha}\left( r,\ell_q \right) &= \int_{\Omega_{\nu_\alpha}}\mathrm d\Omega ~ \! (\hat{\epsilon}^* \lp \hat{q})  \cdot \hat{p} \rp  \\ \nonumber
& =  \int_0^{R_{\nu_\alpha}} \!\!\!\!\!\mathrm dr_d \lp r_d z\rp \Gamma^\perp \lp r_d, \elq, r\rp,
\end{align}
where the dependence on the emission variables is shown. The explicit expression for $\Gamma^\perp$ is given by Eqs.\eqn{e:Gper}-\eqn{e:Gperp}.
Figures \ref{fig:one} and \ref{fig:two} show the geometrical factor \eqn{e:cperp} as a function of the distance travelled by the neutrinos from the neutrinospheres.
The results correspond to the cases A and C (Table \ref{tab:trajectory}) which can be considered as representative of the 
 the typical behaviours of $G^{\perp}_{\nu}$, as we have been observing in our runs.
One can see that $G^{\perp}_{\nu}$ have a similar $r$ dependence as $G_{\nu}$ \eqn{e:G}, as expected. Their absolute values turn out to be
suppressed by a few percent up to several factors, with respect to the $G_{\nu}$ value. 
As we will discuss, the $r$ dependence of  $G_{\nu}$ plays a crucial role on 
the possibility to have multiple crossings and a 
non-linear feedback mechanism
in presence of helicity coherence (see Section IV).
By including Eq.(\ref{e:cperp}) into \eqn{eq:hselfperp} one gets the same expression Eq.\eqn{e:hexpsa} for $h_{\nu \nu}^\perp $
with $G^\perp_{\nu_\alpha}$ Eq.(\ref{e:cperp}) replacing $G_{\nu_\alpha}$ Eq.(\ref{e:geom}).

The neutrino total luminosities and spectra at the neutrinospheres are an essential ingredient of the self-interaction Hamiltonians $h_{\nu \nu} $ and $h_{\nu \nu}^\perp $.
As in Ref.\cite{Frensel:2016fge}, we take the neutrino spectra $f_{\bar{\nu}}$ and $f_{\nu}$  at the neutrinospheres as
Fermi-Dirac distributions,
\begin{equation} \label{Eq:Fermi-Dirac-energy-distribution}
f_{\nu}(p) = \dfrac{1}{F_{2}(0)} \dfrac{1}{T^{3}} \dfrac{p^{2}}{\exp(p / T) + 1},
\end{equation}
where $T $ is the neutrino temperature. In this
expression, we have $F_{2}(0) = \frac{3}{2} \zeta(3)
\approx 1.80$, and $F_{k}(0)$ corresponds to the 
Fermi-Dirac integral of order $k$ with zero degeneracy
parameter,
\begin{equation}
F_{k}(0) \equiv \int_{0}^{\infty} \mathrm{d}x \, \dfrac{x^{k}}{\exp(x) + 1}.
\end{equation}
Table \ref{tab:fluxes} gives the values of the luminosities and average energies for the different neutrino species used in our investigation.

Unoscillated $\nu$ self-interaction potentials constitute a useful quantity to search for the location of helicity coherence resonances. They have been exploited in previous studies of the 
MNR and sMNR Ref.\cite{Malkus:2012ts,Malkus:2014iqa,Zhu:2016mwa,Frensel:2016fge}. 
Such potentials are defined as
\begin{align}\label{e:unpot}
h^{\rm unosc}_{\nu \nu} (r)  & =\sqrt{2} G_F \sum_{\alpha=e,x}
   \int_{0}^{\infty} \! \mathrm dp\ 
 \left[G_{\nu_{\alpha}} (r,\elq) \frac{L_{\nu_{\underline{\alpha}}} f_{\nu_{\underline{\alpha}}}(p)}{ \pi^2 R^2_{\nu_\alpha}\left< E_{\nu_\alpha} \right>} \right. \\
 & \left.-   G_{\bar{\nu}_{\alpha}} (r,\elq)\frac{ L_{ \bar{\nu}_{\underline{\alpha}}} f_{\bar{\nu}_{\underline{\alpha}}}(p)}{ \pi^2 R^2_{\bar{\nu}_\alpha}\left< E_{\bar{\nu}_\alpha} \right>}\right]. \nonumber
\end{align}

\section{Results} 
\subsection{Resonance conditions for helicity coherence}
\noindent
We present here our analysis on the resonance conditions in presence of mass contributions. 
Such situations can be identified by looking at the unoscillated potentials, which will use to characterise our model cases (A, B, C)
that will be presented in Section III.B.
As shown previously, the extended equations with mass terms include both the diagonal Eq.\eqn{eq:Vm}-\eqn{eq:Vm2} and the off-diagonal Eq.\eqn{eq:numix} 
contributions. In the following discussion, we will neglect the diagonal one since they are suppressed by several orders of magnitude
compared to the other terms, as we have been verifying numerically. This is due to the fact that $\vec{V}_m$ Eq.\eqn{eq:Vm} involves correlators with helicity change Eq.\eqn{e:Rgen}, in addition to being proportional to the neutrino mass.

\subsubsection{Majorana case}
\noindent
Conditions for the occurrence of MSW-like resonances\footnote{Note that other resonance phenomena might take place that do not necessarily require such conditions.} are  met when differences of diagonal elements  of the generalized Hamiltonian Eq.\eqn{eq:HacheMajo} become small, i.e. 
$h_{{\cal G},ii} - h_{{\cal G},jj} \simeq 0$  for $i, j =1$ to $4$ ($i \ne j $).

In treatments where neutrino evolution does not include mass terms, neutrino and antineutrino equations of motions are only coupled through the usual self-interaction Hamiltonian Eq.\eqn{e:hexpsa}. In this case, the resonance condition in the neutrino sector reads
\beq\label{e:H12}
h_{{\cal G},11} - h_{{\cal G},22} = - 2 \omega c_{2\theta} + \sqrt{2} G_F n_B Y_e + h_{\nu \nu}^{ee} - h_{\nu \nu}^{xx} \simeq 0.
\eeq
In accretion disks around neutron star merger remnants or black holes, the matter and neutrino self-interaction terms have opposite signs, because  the  
$\bar{\nu}_e$ luminosity is larger than the $\nu_e$ one (see Table \ref{tab:fluxes}). This  can produce a cancellation
of the two contributions. The fulfillment of condition \eqn{e:H12} and the presence of sizeable $H_{ex}$ triggers the MNR resonance phenomenon
where $\nu_e$ change their flavors while $\bar{\nu}_e$ do not.  
The location at which this instability starts can be identified by looking at the matter and unoscillated neutrino profiles, as pointed out in Ref.\cite{Malkus:2014iqa}.
The same cancellation as \eqn{e:H12} can take place in the antineutrino sector, since the resonance condition is given by
\beq\label{e:H34}
h_{{\cal G},33} - h_{{\cal G},44} = -2 \omega c_{2\theta} - \sqrt{2} G_F n_B Y_e - h_{\nu \nu}^{ee} + h_{\nu \nu}^{xx} \simeq 0.
\eeq
Note that depending on the neutrino luminosities and the geometrical factors, the self-interaction term can change sign twice, triggering flavor conversion also in the 
antineutrino sector. This is a necessary condition for the symmetric matter-neutrino resonance  (sMNR) where neutrinos and antineutrinos can modify  their flavors
\cite{Malkus:2012ts}. 

Since we are looking for a situation in which the neutrino-antineutrino coupling produced by the 
$\Phi$ term in Eq.\eqn{eq:HacheMajo} is effective, there are four resonant
conditions between the neutrino and the antineutrino sectors. The first  one is 
\beq\label{e:H13}
h_{{\cal G},11} - h_{{\cal G},33} = \sqrt{2} G_F n_B (3 Y_e  - 1) + 2 h_{\nu \nu}^{ee}  \simeq 0,
\eeq
where we have made use of the explicit expressions for $h_\mathcal{G}$ \eqn{eq:HM}. 
Note that this relation does not involve vacuum terms, and therefore will not depend on the neutrino hierarchy nor on the neutrino energy.
Its fulfillment  involves a cancellation between the matter term and the self-interaction term that is very similar
to the MNR condition \eqn{e:H12}, and it can be identified by using the matter and the unoscillated neutrino self-interaction potential \eqn{e:unpot}.
Relation \eqn{e:H13} can be met if $Y_e > 1/3$ for $h_{\nu \nu}^{ee}  < 0$ or if $Y_e < 1/3$ for $h_{\nu \nu}^{ee}  > 0$.
We remind that here $h_{\nu \nu}$ terms also include trace terms Eqs.\eqn{e:L} and \eqn{e:k}.

The second relation 
\beq\label{e:H14}
h_{{\cal G},11}  - h_{{\cal G},44} = -  2 \omega c_{2\theta} + \sqrt{2} G_F n_B (2 Y_e -1) + h_{\nu \nu}^{ee} + h_{\nu \nu}^{xx} \simeq 0,
\eeq
cannot be satisfied in the standard MNR set-up : a neutron-rich environment which is also $\bar{\nu}_e$ dominated nearby the neutrinosphere with $h_{\nu \nu}^{ee}+ h_{\nu \nu}^{xx}<0$. When a change of sign of $h_{\nu \nu}^{ee}+ h_{\nu \nu}^{xx}$ occurs, which is the case in the sMNR, this resonance may appear.
The third relation 
\beq\label{e:H24}
h_{{\cal G},22} - h_{{\cal G},44} = - \sqrt{2} G_F n_B \lp 1 - Y_e\rp + 2 h_{\nu \nu}^{xx}  \simeq 0.
\eeq
is difficult to meet. Indeed, unless there is a sMNR, $h_{\nu \nu}^{xx}$ is negative, hence \Eqn{e:H24} cannot be fulfilled since $Y_e$ is always smaller than $1$.
Finally the last condition is given by
\beq\label{e:H23}
h_{{\cal G},22} - h_{{\cal G},33} = 2 \omega c_{2\theta} + \sqrt{2} G_F n_B (2 Y_e -1) +  h_{\nu \nu}^{ee}  + h_{\nu \nu}^{xx} \simeq 0,
\eeq
which, as relation \eqn{e:H14}, cannot be met in the case of a standard MNR.
Note that the location of resonances  from the neutrino mass terms are affected by the presence of the MNR resonance, since the MNR
obviously modifies the self-interaction contributions that appear in the helicity resonance conditions.
Note that relations \eqn{e:H13}-\eqn{e:H23} agree with those of Ref.\cite{Cirigliano:2014aoa}.

From \eqn{e:H12}-\eqn{e:H13}, a general relation for the resonance conditions associated with the neutrino mass can be obtained
\begin{align}\label{e:ten}
\sqrt{2} G_F n_B Y_e & >  \sqrt{2} G_F n_B (3 Y_e  - 1) \simeq 2 | h_{\nu \nu}^{ee} |\\
 &  >  | h_{\nu \nu}^{ee} - h_{\nu \nu}^{xx} | > | h_{\nu \nu}^{ee} |. \nonumber
\end{align}
The first inequality holds  if $Y_e < 1/2$, while the second is valid in the case of a standard MNR, where $\left| h_{\nu \nu}^{ee} \right| >  \left| h_{\nu \nu}^{xx} \right| $.
The central approximate equality corresponds to relation \eqn{e:H13}, while the two quantities on the left and on the right correspond to the
MNR resonance condition \eqn{e:H12}. Relation \eqn{e:ten} shows that the standard MNR and the helicity coherence condition \eqn{e:H13} cannot be satisfied simultaneously, while this is possible in the case of a symmetric MNR.

 \subsubsection{Dirac case}
\noindent
If neutrinos are Dirac particles, the generalized Hamiltonian that governs the evolution is given by
Eq.\eqn{eq:HD}. In this case the resonance conditions read
\beq\label{e:rD1}
h_{D,{\cal G},11} - h_{D,{\cal G},33} = \frac{1}{2} \left[ h_{{\cal G},11} - h_{{\cal G},33} \right] \simeq 0,
\eeq
\beq\label{e:rD2}
h_{D,{\cal G},22} - h_{D,{\cal G},44} =\frac{1}{2} \left[  h_{{\cal G},22} - h_{{\cal G},44} \right]  \simeq 0,
\eeq
\beq\label{e:rD3}
h_{D,{\cal G},11} - h_{D,{\cal G},44} = \frac{1}{2} \left[ h_{{\cal G},11} - h_{{\cal G},33}   \right] -2 \omega c_{\theta}\simeq 0,
\eeq
\beq\label{e:rD4}
h_{D,{\cal G},22} - h_{D,{\cal G},33} = \frac{1}{2} \left[ h_{{\cal G},22} - h_{{\cal G},44}\right] + 2 \omega c_{\theta} \simeq 0.
\eeq
In the Dirac case the two conditions Eqs.\eqn{e:rD1} and \eqn{e:rD3} can be satisfied in the same conditions than \Eqn{e:H13}; while
Eqs.\eqn{e:rD2} and \eqn{e:rD4} requires a change of sign of  $h_{\nu \nu}^{ee}+ h_{\nu \nu}^{xx}$.

\subsection{Numerical results on flavor evolution}

\noindent
We now present our numerical results on flavor evolution. We show neutrino survival probabilities and quantify the adiabaticity of neutrino evolution through the resonances. We have studied a large ensemble of conditions,
both for the potential profiles and parameters. 
We emphasize that computations are particularly demanding : indeed, we solve the coupled evolution equations of the full $4\times 4$ generalized density matrices with four different initial conditions, in a two-dimensional model, using $10^3$ energy bins. We present results on the neutrino evolution up to $300$ km from the neutrinosphere, distance at which the numerical convergence is achieved. 
Note that the inputs from BNS merger simulations Ref.\cite{Perego:2014fma} have been obtained following the same procedure
as in Ref.\cite{Frensel:2016fge}. 
\begin{table} 
\center
\begin{tabular}{ccccc}
Model &  Type & $x_0$ & $z_0$ & $\theta_0$  \hfill \\ \hline
A  &  MNR  & 15 & 32 & $15^{\circ}$ \\
B  &  helicity coherence   &12  & 27 & $40^{\circ}$\\
C  &  MNR and helicity coherence & -30 & 20 & $55^{\circ}$ \\
 \hline
\end{tabular}
\caption{Characteristics of the three scenarios considered in our schematic model. 
The second and third columns give the location of the neutrino emission point $x_0$ (km), $z_0$ (km)  while $\theta_q$ defines the neutrino trajectory in the $\pi_{xz}$ plane (Figure \ref{fig:schem}).}  
\label{tab:trajectory}
\end{table}

In the present study, we have searched mostly for the helicity coherence resonance conditions \eqn{e:H13}, which is the most interesting one
in our astrophysical setting, as well as more generally when $ Y_e < 1/2$\footnote{We are interested in flavor modifications in relation with nucleo-synthesis where neutrinos plays a role, such as the $r$-process.}.
We choose to present three model cases A, B, and C, that correspond to different astrophysical conditions during neutrino evolution. 
In Model A the MNR condition \Eqn{e:H12} is met, while mass contributions are included without fulfillment of the helicity coherence resonance \Eqn{e:H13}. 
In Model B, the helicity coherence resonance condition Eq.\eqn{e:H13} is met, while the MNR condition, which is also met, leads to no flavor conversions.
Model C has both the MNR \eqn{e:H12} and  helicity \eqn{e:H13} conditions fulfilled and the MNR effectively leads to flavor conversions.
Table \ref{tab:trajectory} shows the initial location and the angles defining
the neutrino trajectory followed in the three models.
Note that we set  $\phi_q=0$ since neutrinos follow straight line trajectories.

In order to fully unravel the effects of the mass terms we have explored for each parameter a range of values.
For the total neutrino luminosity, we have used values from Ref.\cite{Perego:2014fma} and rescaled ones,
to investigate luminosity variations within the range compatible with available BNS merger simulations (see Ref.\cite{Frensel:2016fge} for a detailed discussion). For the anisotropic matter term, we have considered matter velocities in the range $\beta \in [0.05, 0.1]$, the value of $\beta=0.1$ being an upper bound for this type of scenarios. 
In particular, we make the {\it ansatz} that the perpendicular quantity is of the same order as the radial ones (see Figures 15, 16, 19 of Ref.\cite{Perego:2014fma}).
Our numerical results show that anisotropy  from the matter currents is always suppressed  compared to the neutrino current anisotropy. Therefore our optimistic {\it ansatz} for the perpendicular velocities will have little impact on our conclusions. The results shown below are all obtained with the value $\beta = 0.1$.

The additional contributions due to the neutrino mass depend on the mass matrix Eq.(\ref{e:Mdec}).
The neutrino mixing parameters used in our simulations are
$\Delta m^2  = 2.43 \times 10^{-3} $ eV$^2$ and $\sin^2 \theta = 0.087$, consistently with measured
values \cite{Agashe:2014kda}. As for the hierarchy, which is still unknown, the mass effects do not appear to depend on the sign of
$\Delta m^2 $,  Eq.\eqn{e:H13}.
A slight dependence is present when the MNR occurs.  
We have performed calculations both by taking/neglecting the $\Delta m^2$ term in Eq.(\ref{e:Mdec}) and the Majorana phase.
Our results turn out to be insensitive to them.

Adiabaticity of the evolution at a resonance location is crucial for flavor or helicity conversions to occur.
Different approaches can be used to quantify it (see e.g. \cite{Duan:2010bg,Galais:2011jh}), including the ${\rm SU}(2)$ neutrino isospin formalism which is applicable to the two-flavor framework  (Appendix C). Since, in our model, two neutrino and antineutrino flavors are coupled to each other, the density matrix is a $ 4 \times 4$ matrix. However, it turns out that in most cases,  
either there are flavor conversions because of the MNR while neutrinos and antineutrinos propagations are decoupled; or
 the helicity coherence resonance is met while MNR is ineffective. Therefore, we can effectively apply the ${\rm SU}(2)$ neutrino isospin formalism to our system. In the numerical results presented below, the angle between the effective isospin and magnetic field will be shown to quantify adiabaticity.

\subsubsection{Model A}
\label{subsub:modelA}

\begin{figure}[htpb]
\begin{center}
\includegraphics[scale=0.377]{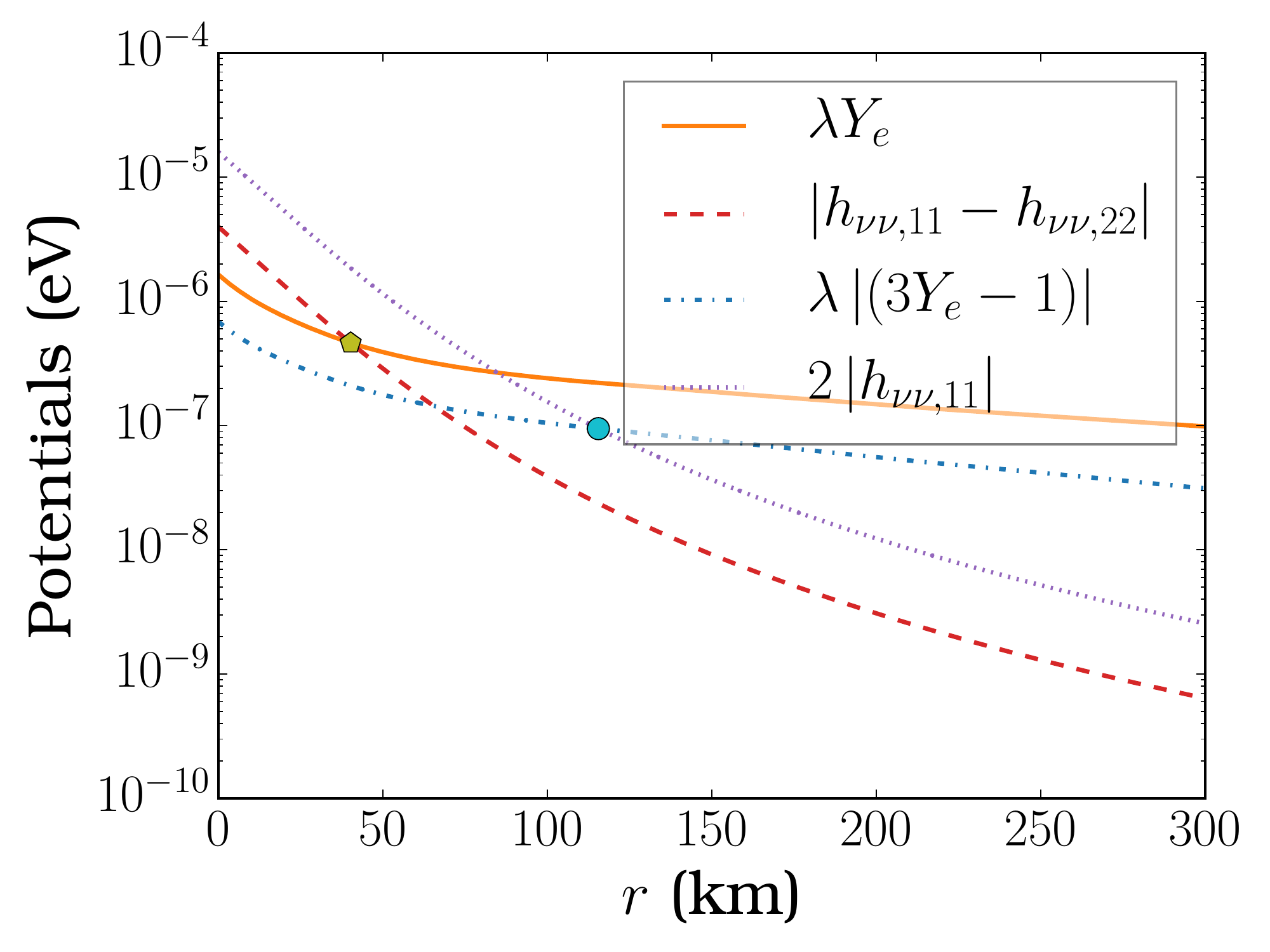}
\caption{Matrix elements appearing in the MNR Eq.\eqn{e:H12} and helicity resonance conditions \eqn{e:H13} for Model A (Table \ref{tab:trajectory}).
The values correspond to the matter and to the unoscillated self-interaction potentials Eqs.\eqn{e:hmat2f} and \eqn{e:unpot} respectively. The green pentagon shows the location of the beginning of the MNR, while the blue dot shows the location of the helicity coherence resonance.}
\label{fig1}
\end{center}
\end{figure}

In this first model, our goal is to establish whether some effects due to neutrino mass would appear in the absence of a helicity coherence resonance. For this reference case, the luminosities used are rescaled $L_{\nu_e, \text{res}} = 0.65 L_{\nu_e}$, $L_{\nu_x, \text{res}} = 1.16 L_{\nu_x}$, while the $\bar{\nu}_e$ luminosity is unchanged.
Figure \ref{fig1} shows the matter and unoscillated $\nu$-$\nu$ potential Eq.\eqn{e:unpot} for Model A.
While neutrino self-interaction is larger than  the matter potential  close to the neutrinosphere, they
cross at $40$ km, the location for a MNR resonance. 
In Model A, though there is a helicity coherence resonance which would occur around $150$ km due to flavor conversions, 
we will focus on the region before to show a reference calculation where the resonance helicity condition is not fulfilled.
We expect that, in the absence of resonance condition for the mass terms, no new effects appear in the MNR region, since the coupling $\frac{m}{q}$ is small. Indeed, we find the same flavor conversion due to the MNR, 
as in absence of mass contributions. 

\begin{figure}[!thb]
    \centering
    \begin{minipage}{.5\textwidth}
        \centering
        \includegraphics[scale=0.377]{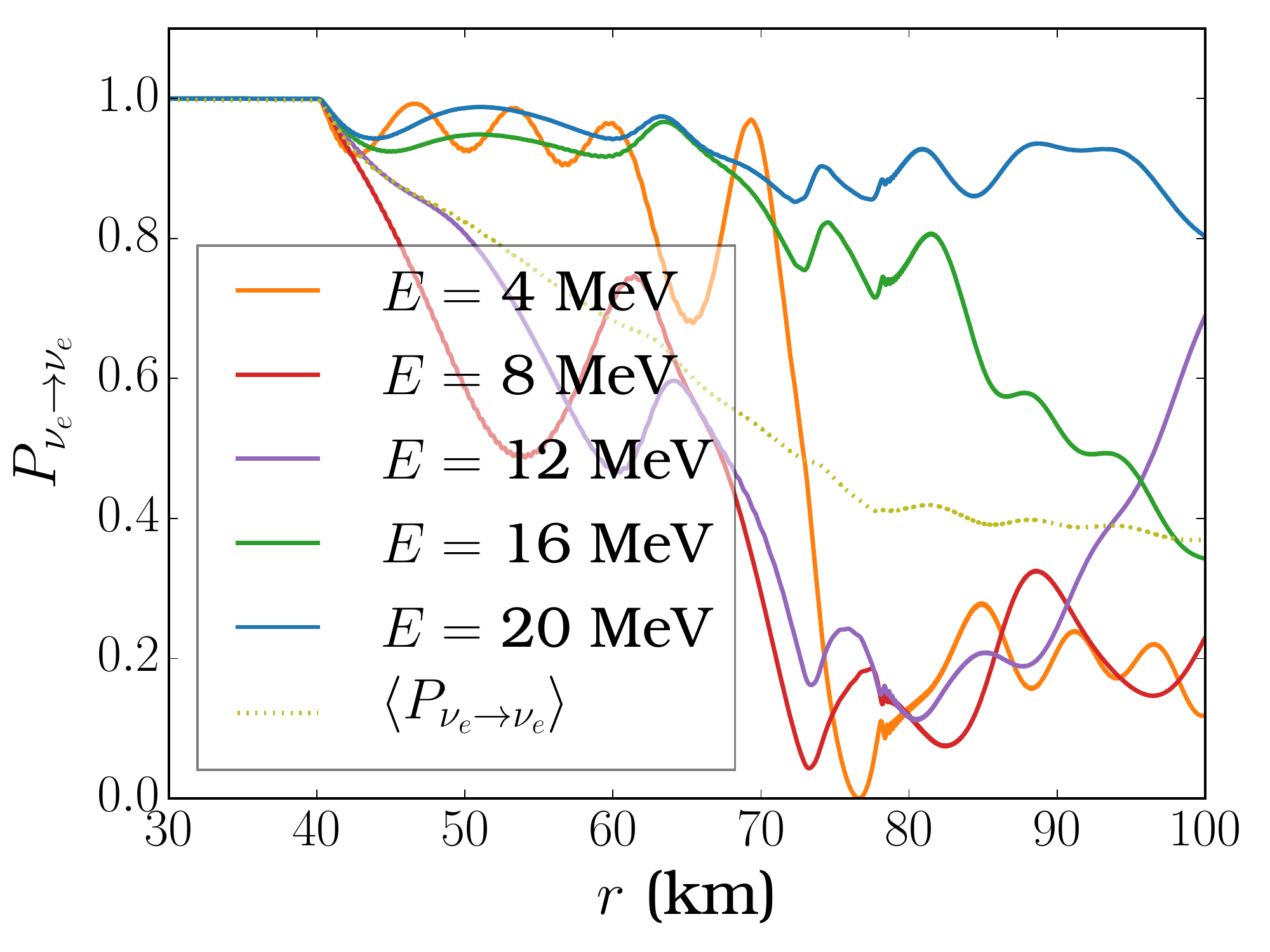}
        \includegraphics[scale=0.377]{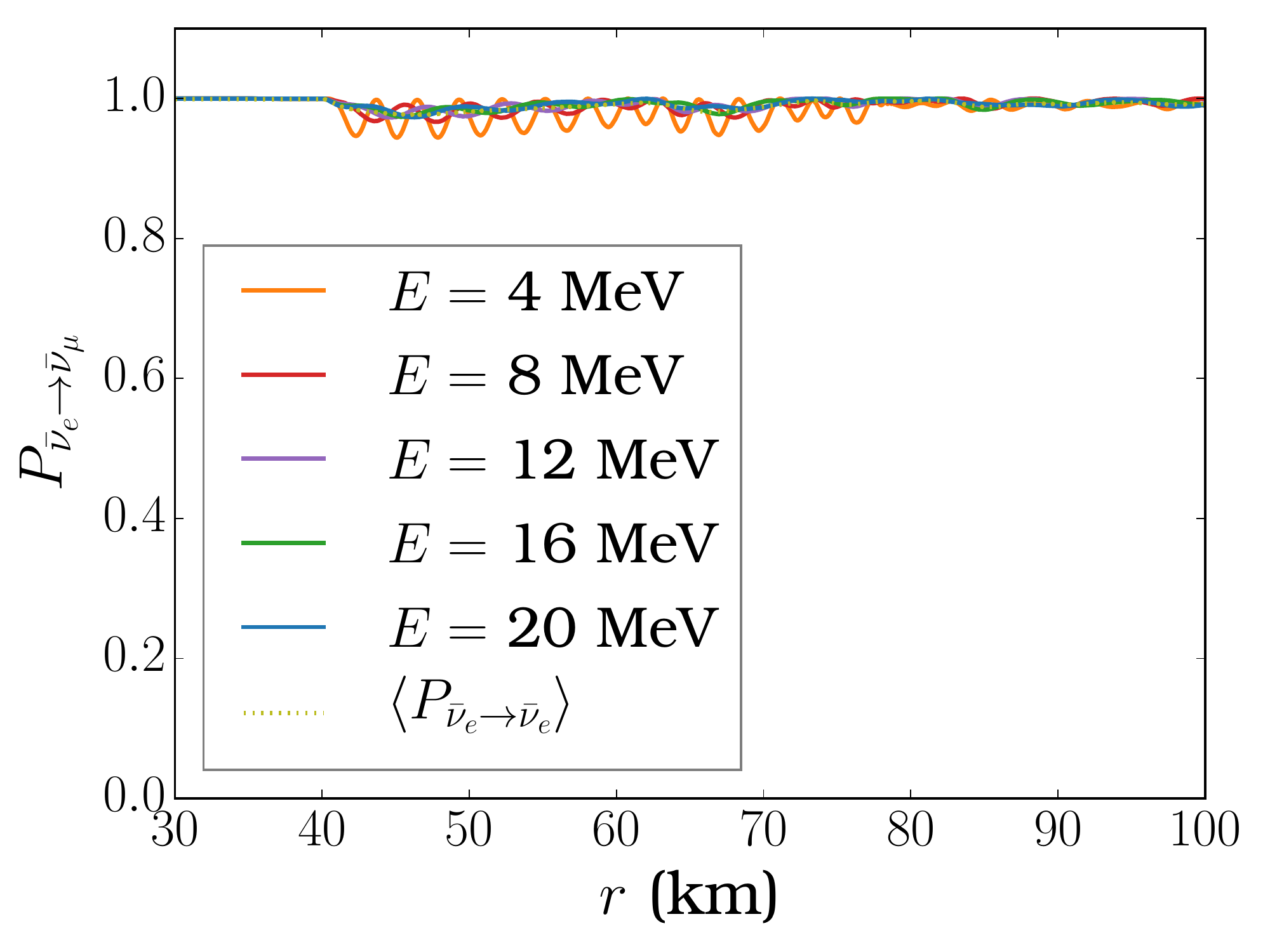}
        \includegraphics[scale=0.377]{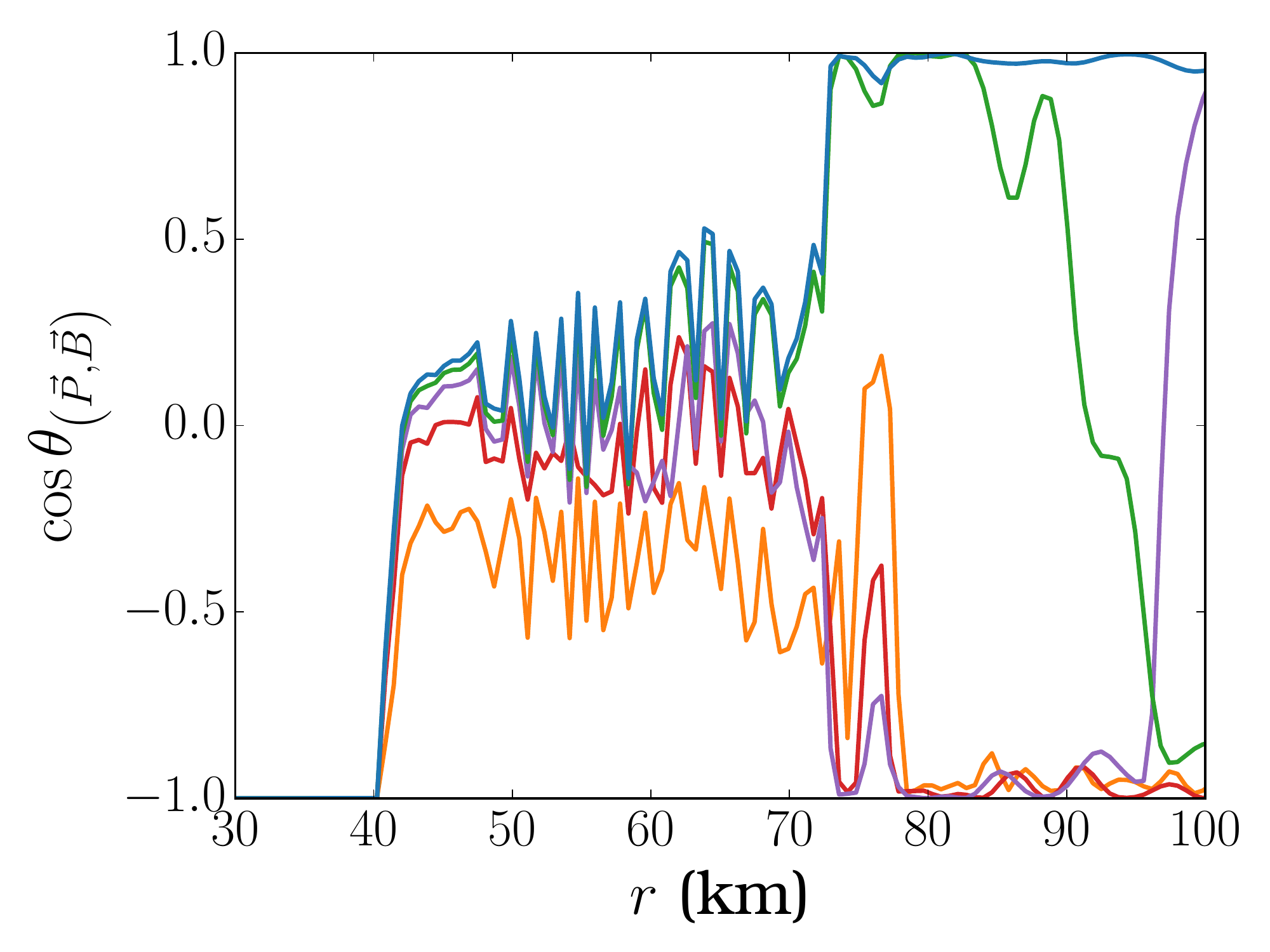}
        \caption{Model A : Electron neutrino (upper) and antineutrino (middle figure) survival probabilities for different energies,  in presence of a MNR starting at $40$ km (see Figure \ref{fig1}). The averaged probability is also presented. The lower figure shows the locally-averaged cosine of the angle between the effective isospin and magnetic field. }
\label{fig:PmodA}
    \end{minipage}
\end{figure}

The numerical results here are given for $\alpha=\frac{5\pi}{6}$ and $m_0=0.1$ eV. The survival probabilities for neutrinos and antineutrinos are given in Figure \ref{fig:PmodA} for the MNR resonance region only and for several neutrino energies. One can see that electron neutrinos efficiently convert into $\nu_x$ whereas the antineutrinos do not modify their flavor content, which is characteristic of MNR.

For neutrinos, there is an energy range between $4$ MeV and $13$ MeV for which flavor conversions are efficient. Above $13$ MeV, though the resonance condition is fulfilled, 
Figure \ref{fig:PmodA} shows that the isospins do not follow the evolution of the effective magnetic field, making the resonance non-adiabatic.
A detailed discussion on adiabacity in presence of the MNR is made within schematic models in Refs.\cite{Wu:2015fga,Vaananen:2015hfa}.
In order to establish the importance of each term in \Eqn{e:H12} to maintain the resonance over such a long distance, we have performed a run where artificially the oscillating part of the term $h_{\nu\nu}^{xx}$ is set to zero (keeping only the trace part). The results are intriguing, since we find that even with this term set to zero, the resonance still maintains over tens of kilometers, the value of $h_{\nu\nu}^{ee}$ being readjusted at each time by the non-linearity.

\subsubsection{Model B}

\begin{figure}[htpb]
\begin{center}
\includegraphics[scale=0.377]{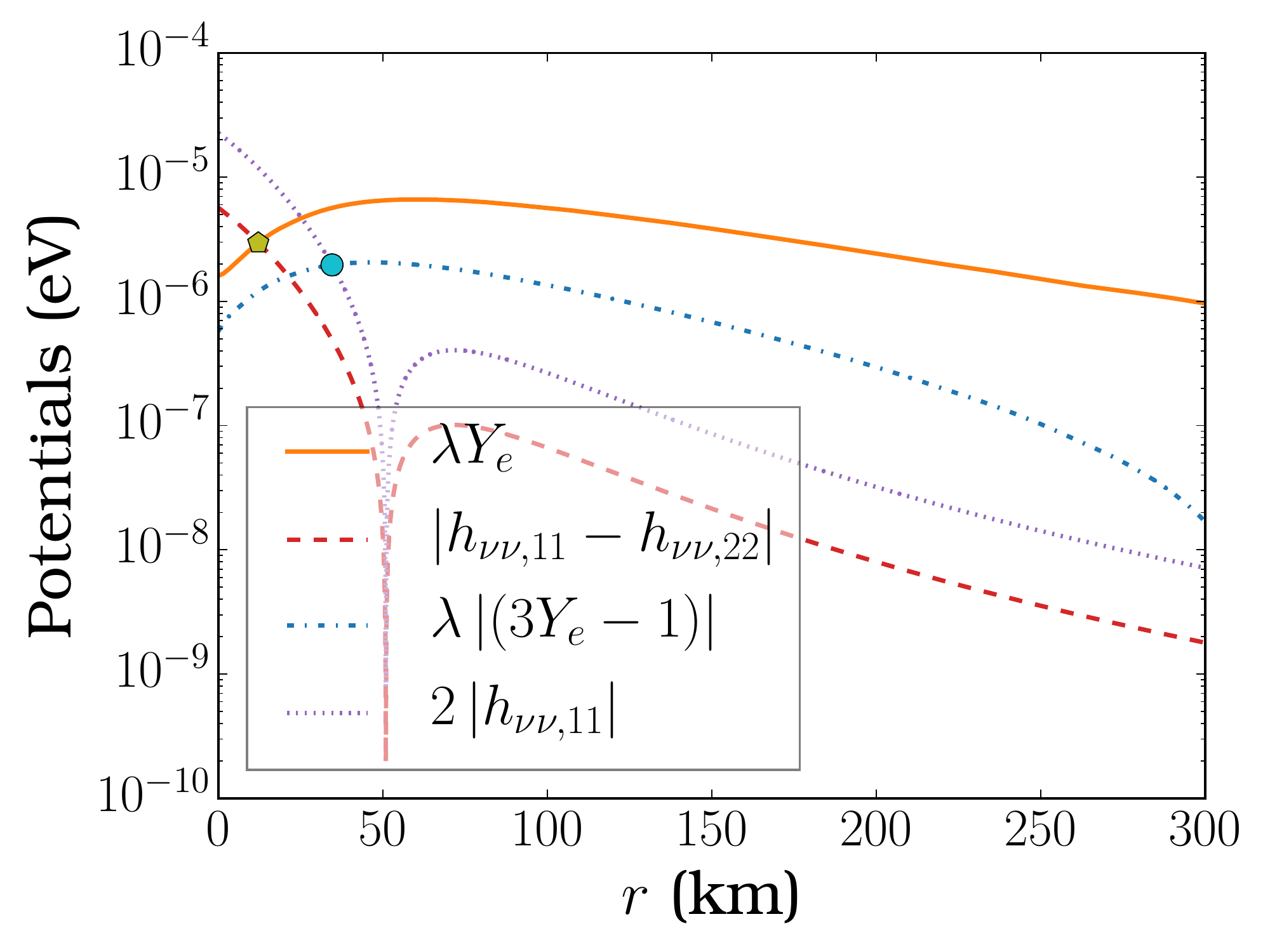}
\caption{Same as Figure \ref{fig1} but for Model B (Table \ref{tab:trajectory}).}
\label{fig2}
\end{center}
\end{figure}

Having shown that in the absence of helicity coherence resonance, no effects arise from the mass terms, we now explore the case in which there are resonances. Results for the matter and unoscillated $\nu$-$\nu$ potential Eq.\eqn{e:unpot} for Model B are shown in Figure \ref{fig2}. Nearby the neutrinosphere, the neutrino potential dominates over the matter one, while after a few tens of km the situation gets
reversed : the MNR condition is met at the crossing point, around $12$ km. However, the adiabaticity of the evolution is not sufficient to trigger flavor conversions.
On the other hand, the helicity coherence resonance Eq.\eqn{e:H13} is met at $34$ km.

As explained before, the computations in this scenario are very demanding. Since we established that in the absence of a helicity coherence resonance, the results were the same for the full $4\times4$ problem as for two decoupled $2\times2$ neutrino and antineutrino matrices, 
we solve the full problem around the helicity coherence resonances using as initial conditions the results obtained in the absence of the mass couplings\footnote{This is done  to keep the computational times manageable.}. Note that the results correspond to the first helicity coherence resonance in Figure \ref{fig2}. Similar results were obtained for the second resonance.

\begin{figure}[!thb]
    \centering
    \begin{minipage}{.5\textwidth}
        \centering
        \includegraphics[scale=0.37]{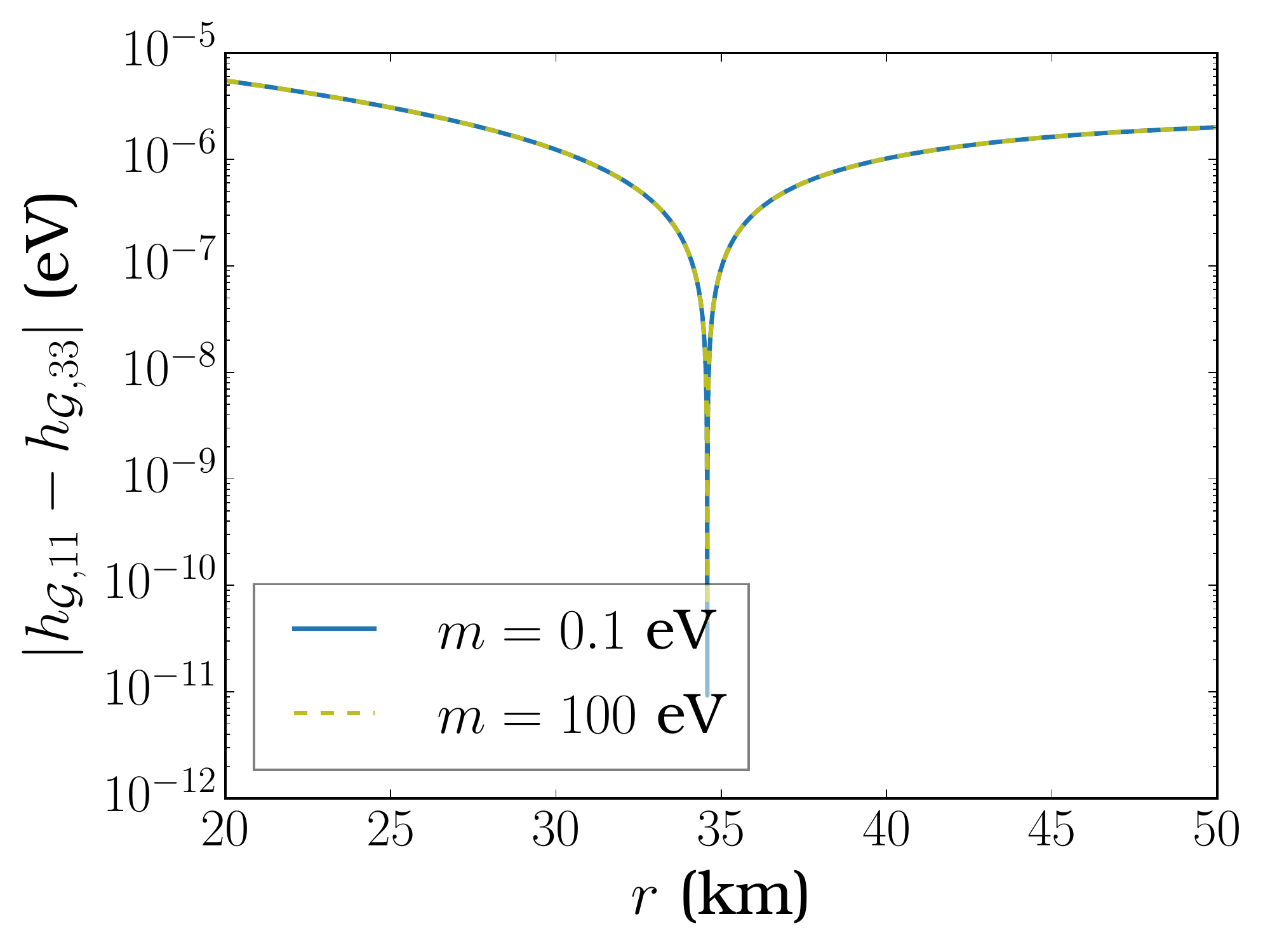}
        \includegraphics[scale=0.37]{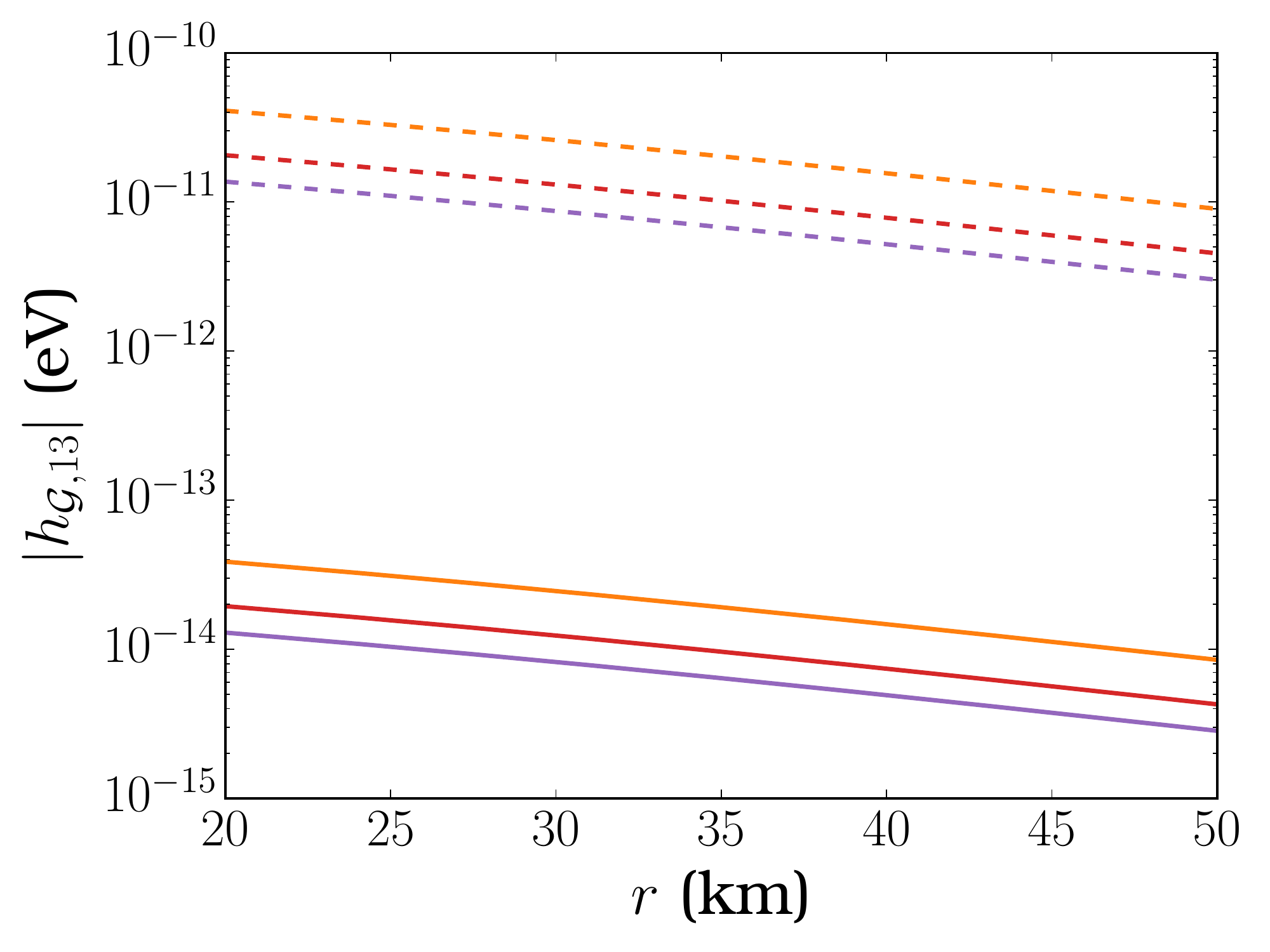}
        \caption{Model B : Resonance condition \eqn{e:H13} (upper) as well as the off-diagonal matrix element 
$h_{{\cal G},13}$ (lower figure) that is non-zero in presence of the neutrino mass. The helicity coherence resonance can be seen
at $34$ km. The different colors correspond to different neutrino energies. The solid lines are the results for $m_0=0.1$ eV, while the dotted lines are for the unrealistic value of $m_0=100$ eV. }
\label{fig:HModB}
    \end{minipage}
\end{figure}

Figure \ref{fig:HModB} shows the resonance condition \eqn{e:H13} as well as the off-diagonal matrix element
$h_{{\cal G},13}$  that is non-zero in presence of the neutrino mass, for an absolute mass of $m_0=0.1$ eV, and for the unrealistic value $m_0 =100$ eV. In both cases, the Majorana phase is taken to be $\alpha=\frac{\pi}{3}$. This case is taken as an example to point out that, even when the off-diagonal terms are multiplied by a factor of $10^3$
artificially, it is not sufficient to trigger a non-linear feedback mechanism and the resonance width stays very narrow. We will elaborate on this aspect in Section \ref{sec:nlf}.

\begin{figure}[!thb]
    \centering
    \begin{minipage}{.5\textwidth}
        \centering
        \includegraphics[scale=0.37]{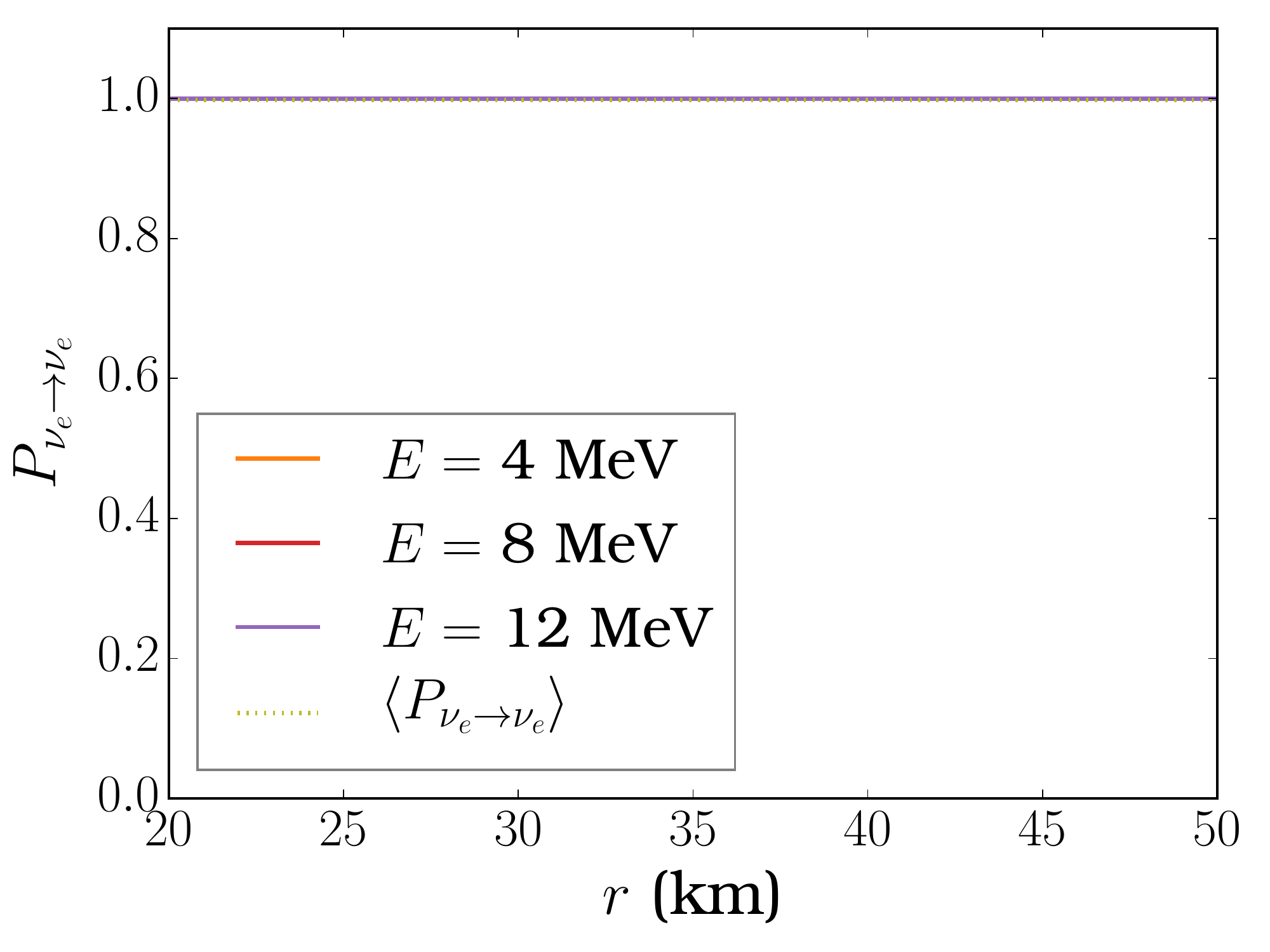}
       \hspace*{-.5cm}
        \includegraphics[scale=0.395]{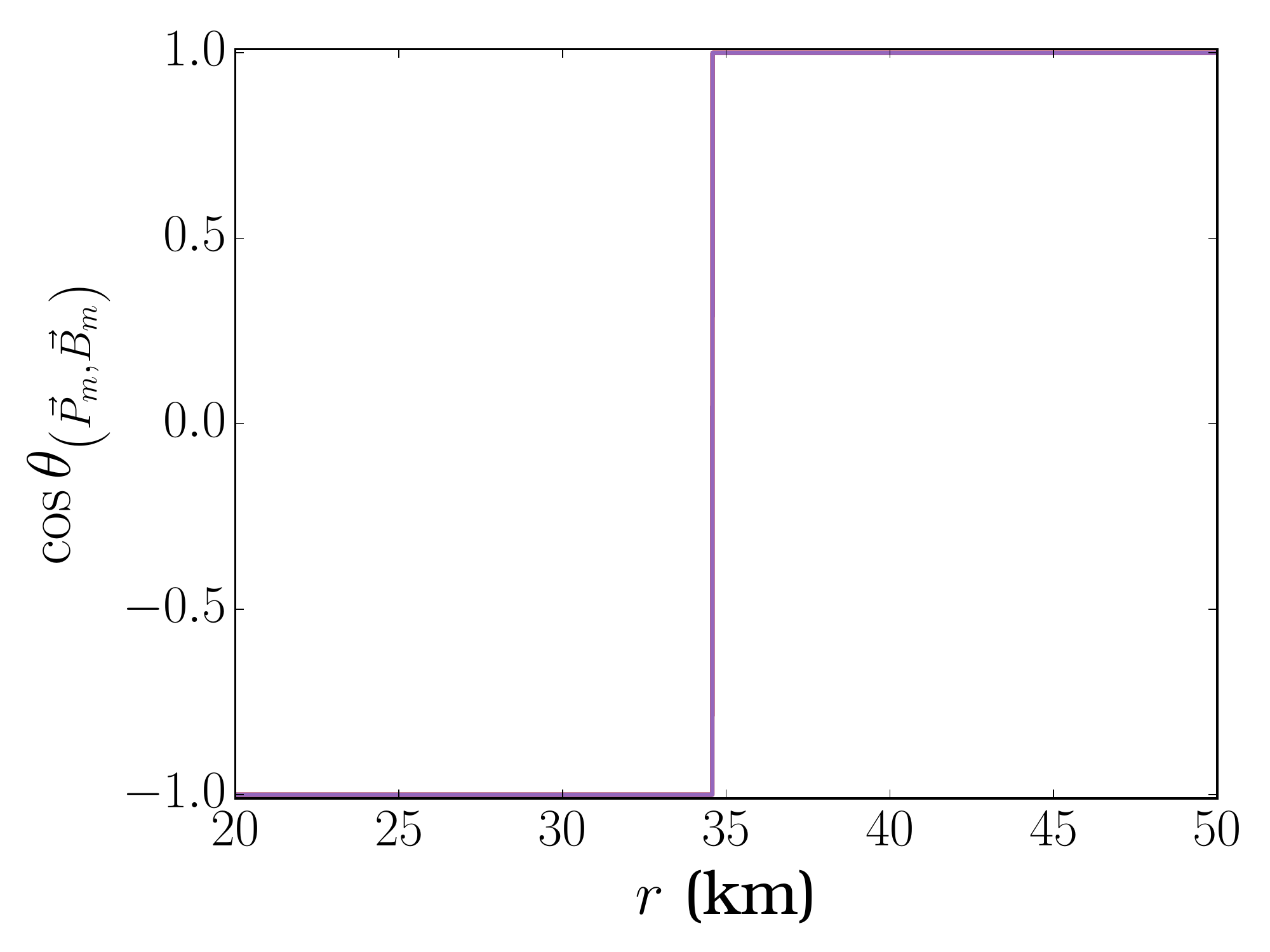}
        \caption{Model B : Electron neutrino survival probability (upper) and adiabaticity (lower figures). The results for different energies are indistinguishable.}
\label{fig:PmodB}
    \end{minipage}
\end{figure}

Figure \ref{fig:PmodB} shows the electron neutrino survival probability and the angle quantifying the adiabaticity through the  helicity coherence resonance for three different energies as typical examples. As one can see the evolution is completely non-adiabatic at the resonance, explaining why there is no helicity conversions. Note that the evolution stays non-adiabatic even when the absolute neutrino mass is larger by a factor of
$10^{3}$.

\subsubsection{Model C}

\begin{figure}
\begin{center}
\includegraphics[scale=0.4]{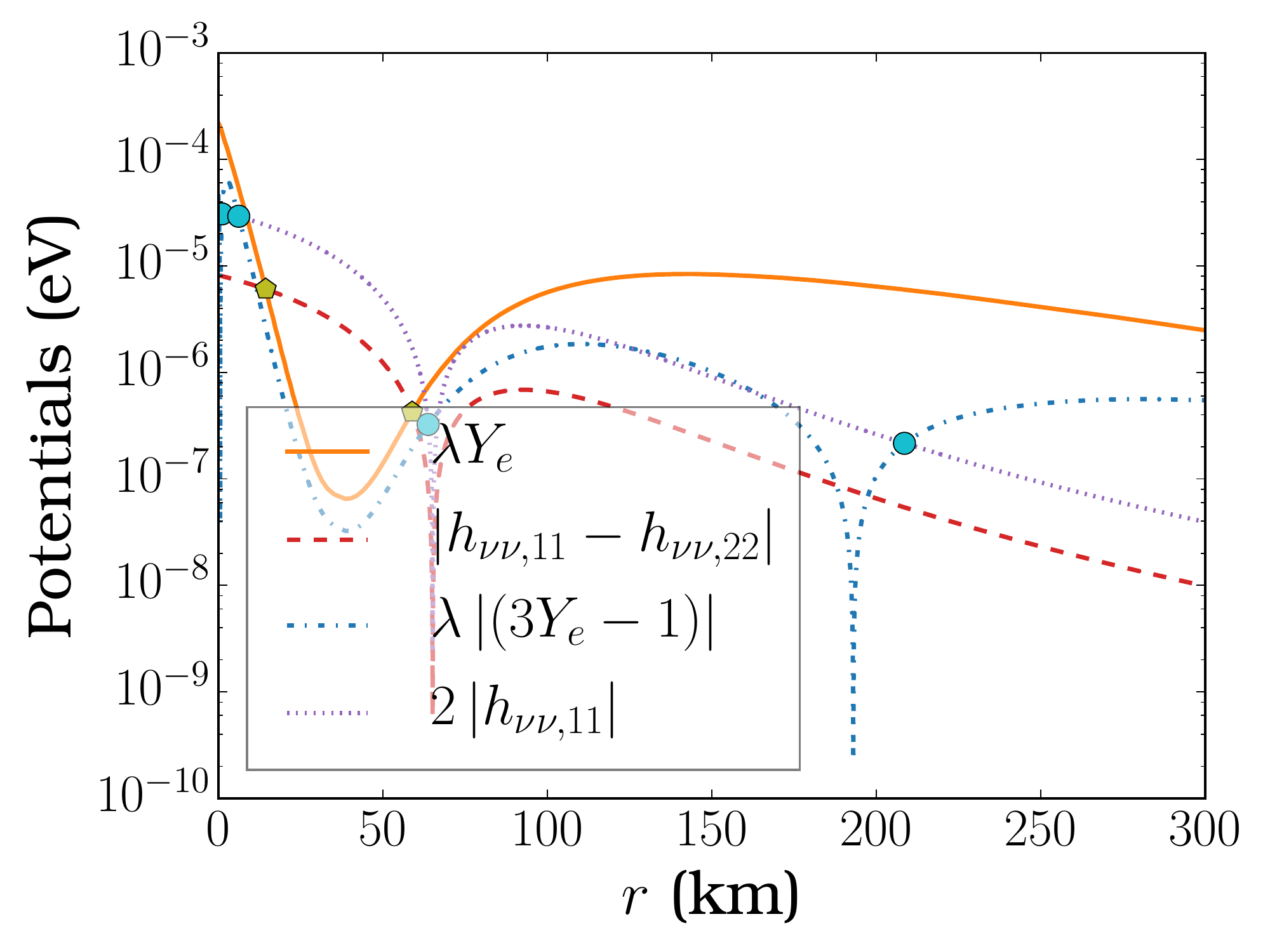}
\caption{Same as Figure \ref{fig1} but for Model C (Table \ref{tab:trajectory}).}
\label{fig3}
\end{center}
\end{figure}

In model C, we look into the scenario where both an effective MNR and the helicity coherence resonance are met. The luminosities used here are 
rescaled $\nu_e$ and $\bar{\nu}_e$ luminosities $L_{\nu_e, \text{res}} = 1.67 L_{\nu_e}$, $L_{\bar{\nu}_e, \text{res}} = 1.1 L_{\bar{\nu}_e}$, while the $\nu_x$ luminosities are unchanged.  Figure \ref{fig3} shows the matter potentials and the unoscillated neutrino potentials. In the first kilometers, the matter dominates over the neutrino potential, with two helicity coherence resonances around $2$ km and $7$ km, up to $15$ km where a first MNR crossing occurs. Then, the neutrino potential dominates until the second MNR crossing at $59$ km, which is a symmetric MNR. There is another helicity coherence resonance at $64$ km. In this model, there is a change of sign for $h_{\nu\nu}^{ee}$, and a little bit later, $Y_e$ goes from $Y_e>\frac{1}{3}$ to $Y_e<\frac{1}{3}$ : because of these two changes, there is a fourth helicity coherence resonance at $208$ km.

The first two helicity coherence resonances are very similar to the one observed in model B, because they occur prior to any flavor conversions. Indeed, numerical computations give the same results as before : a very narrow resonance, without any helicity conversion. The first MNR crossing does not lead to any flavor conversions, the adiabaticity of the evolution being not sufficient, while the second MNR crossing is efficient. Because of these conversions, the potential $h_{\nu\nu}^{ee}$ is modified and no longer changes of sign. The oscillated neutrino potentials obtained with our $2\times 2$ code shows that because of flavor conversions, 
the last two helicity coherence resonances are turned into three resonances at $70$, $82$ and $91$ km.

\begin{figure}
\begin{center}
       \includegraphics[scale=0.37]{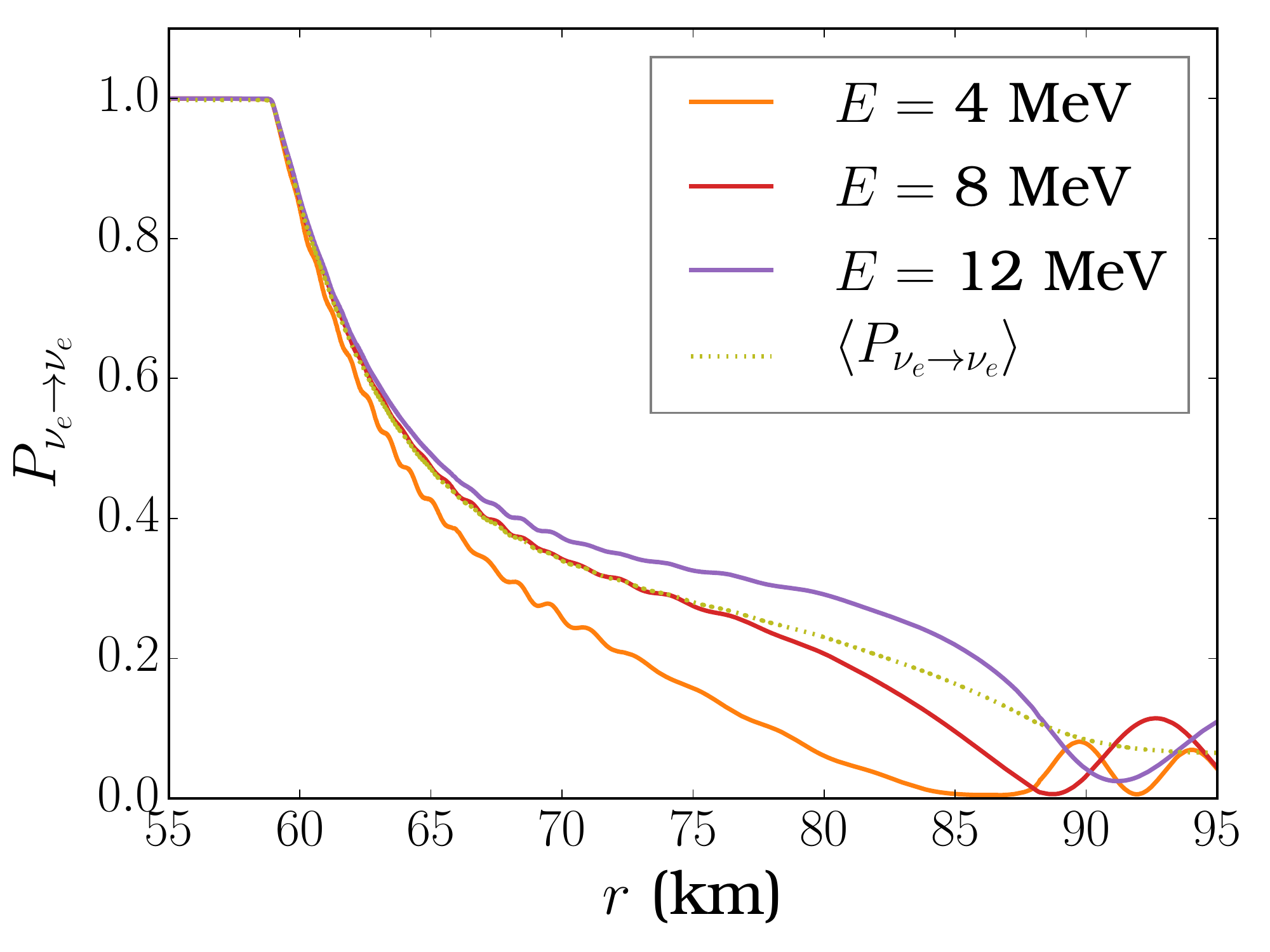}
       \includegraphics[scale=0.37]{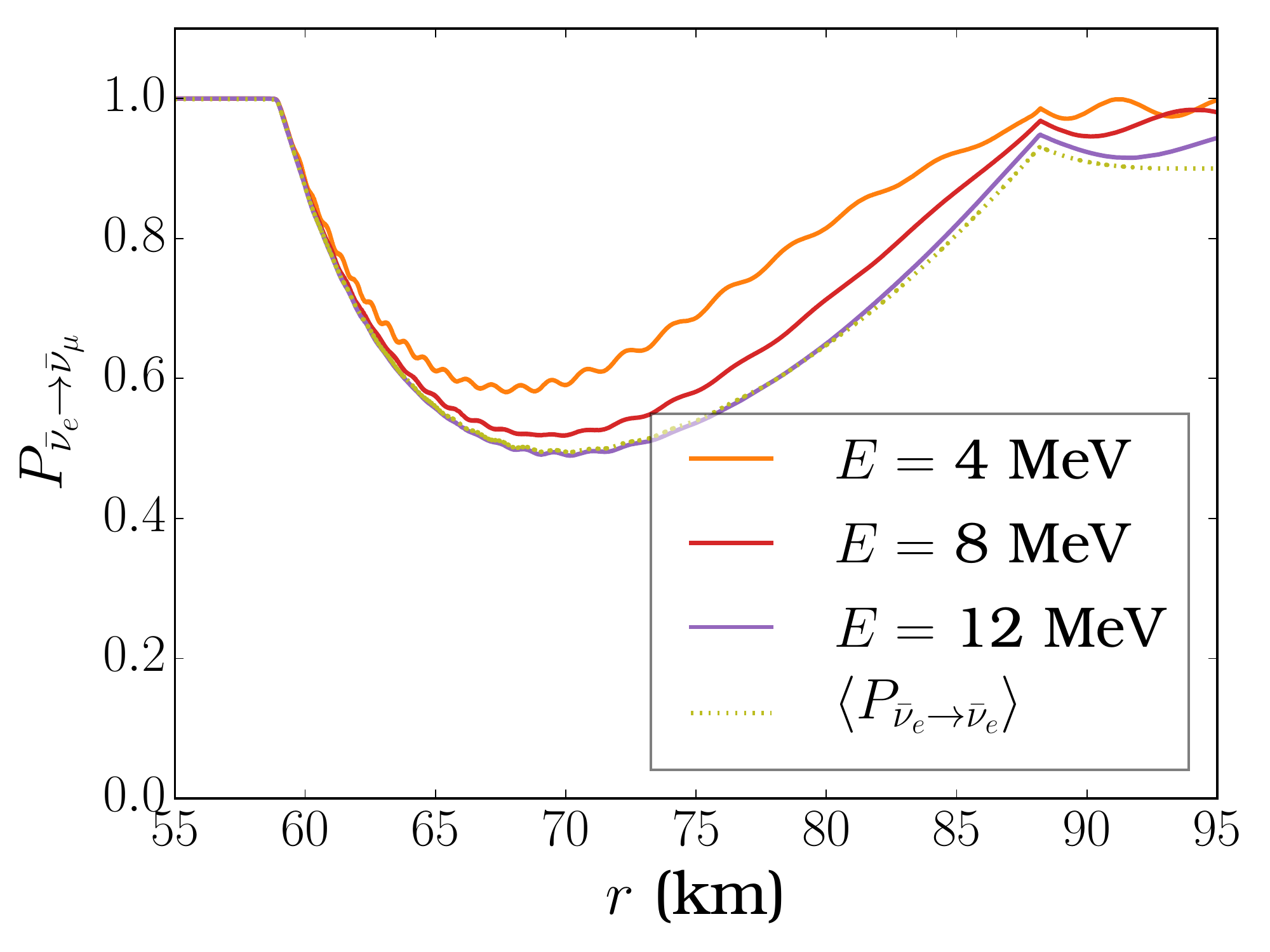}

        \caption{Model C : Electron neutrino (upper) and antineutrino (lower figure) survival probabilities for different energies, in presence of a symmetric MNR at $59$ km (see Figure \ref{fig3}) and of a helicity coherence resonance around $103$ km. The averaged probabilities are also shown.}
\label{fig4}
\end{center}
\end{figure}

We numerically investigated these resonances, which are superimposed with the symmetric MNR, and obtained the same results as for the symmetric MNR without mass terms. Neutrinos and anti-neutrinos survival probabilities are shown in Figure \ref{fig4} for different neutrino energies, in the region where both the MNR and the helicity resonance condition are fulfilled. At the MNR, neutrinos undergo a strong (adiabatic) conversion while antineutrinos evolve semi-adiabatically through the resonance. At the helicity coherence resonance, both neutrinos and antineutrinos have a non-adiabatic evolution.

\begin{figure}
\begin{center}
         \includegraphics[scale=0.37]{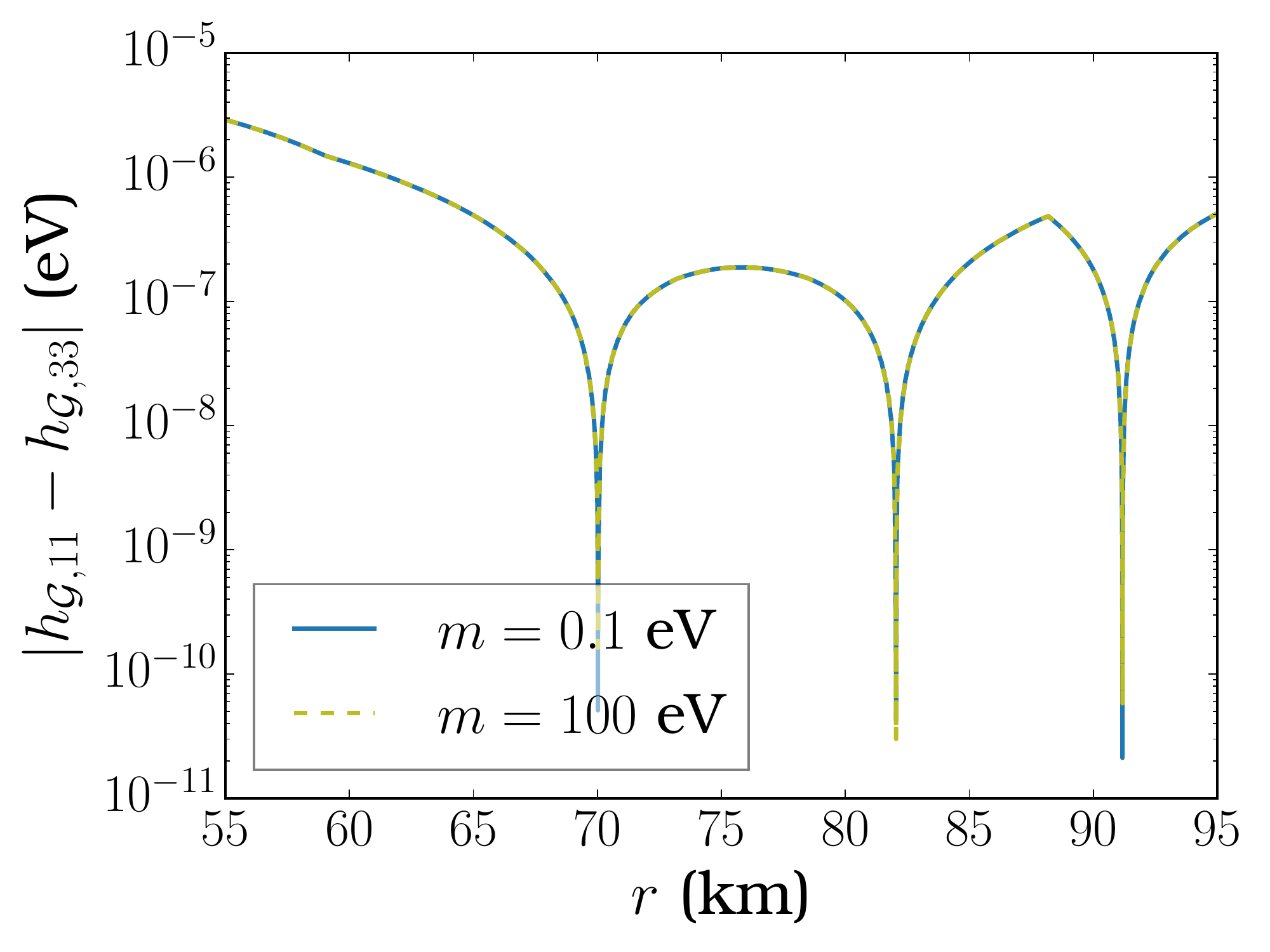}
        \includegraphics[scale=0.37]{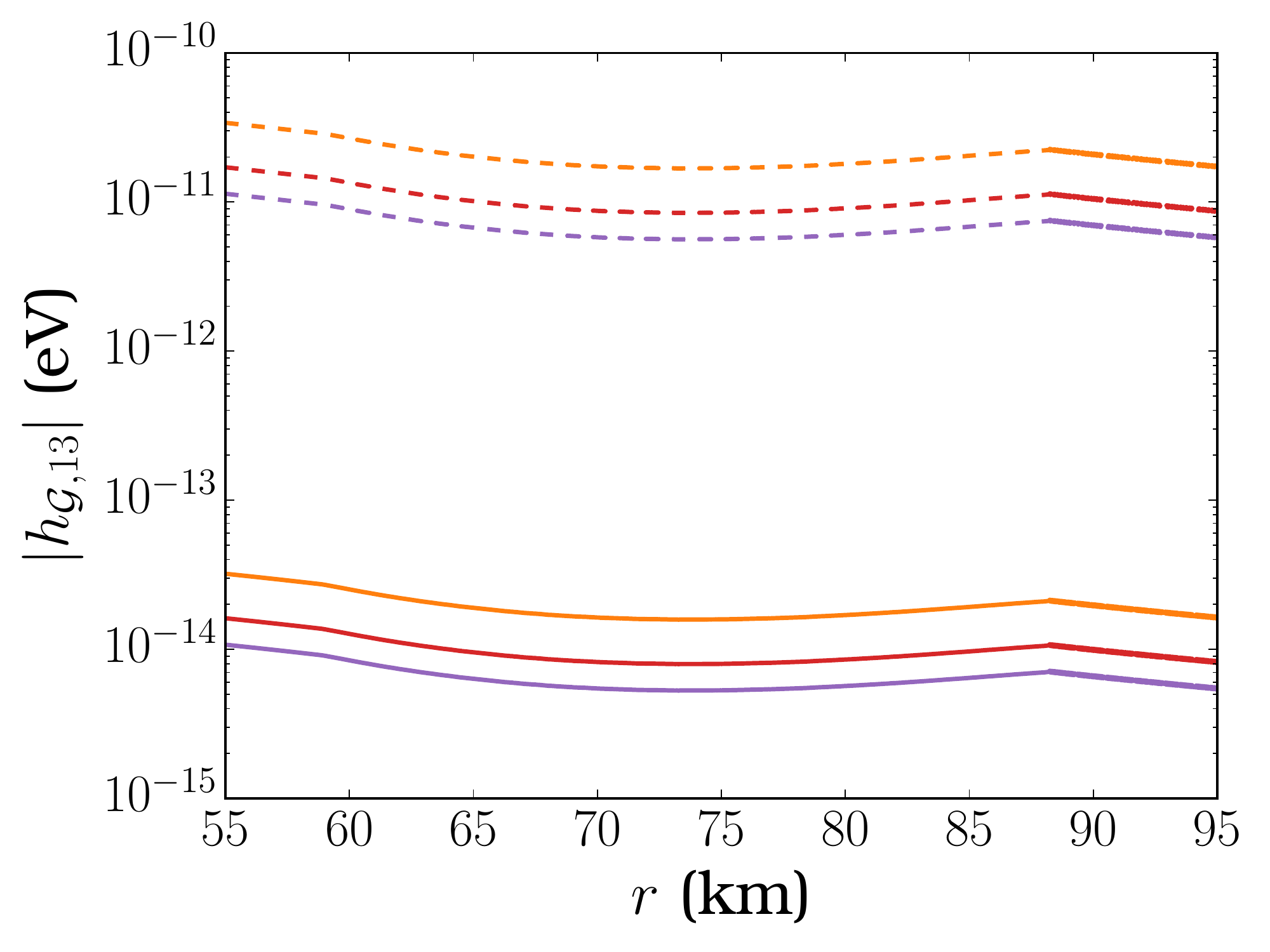}
        \caption{Model C : Resonance condition \eqn{e:H13} and the off-diagonal matrix element
$h_{{\cal G},13}$ for two different value of the absolute mass $m_0$.}
\label{fig:h13C}
\end{center}
\end{figure}

Figure \ref{fig:h13C} shows $| h_{\mathcal{G},11} - h_{\mathcal{G},33} |$ and its associated off-diagonal element with two different values of the neutrino absolute mass $m_0=0.1$ eV and $m_0=100$ eV, around the helicity coherence resonance. As in the case of model B, we take  $\alpha=\frac{\pi}{3}$. Their behaviors are similar to those of model B. In particular, despite having 
$|h_{\mathcal{G},13}|$ close to  $| h_{\mathcal{G},11} - h_{\mathcal{G},33} |$ for the lower energies and for $m_0 = 100 $ eV, the resonance is too narrow to render helicity conversions possible.

Note that, in this model, the MNR is symmetric, and  $h_{\nu \nu}^{ee}+ h_{\nu \nu}^{xx}$ changes of sign. Because of this, we also meet the three other resonances Eqs.\eqn{e:H14}, \eqn{e:H23} and \eqn{e:H24}. Numerical investigations show that they are very similar to the helicity coherence resonance \Eqn{e:H13} : the evolution through these extremely narrow resonances is completely non-adiabatic, hence, no conversion occurs.

\section{Non-linear feedback mechanisms} 
\label{sec:nlf}
We discuss here general aspects of the conditions to have multiple MSW resonances and a non-linear feedback mechanism. By using first-order perturbative developments of the matrix elements, we first analyse two cases where such mechanisms operate, using heuristic arguments. Then, we study why the necessary matching conditions are difficult to meet in more realistic helicity coherence models. 
Obviously the arguments we give are valid if the average variations on short time scales catch the behaviour on larger time scales. 

\subsection{Non-linear feedback in the MNR}
\label{sub:nlfeedMNR}
\begin{figure}[htpb]
\begin{center}
\includegraphics[scale=0.4]{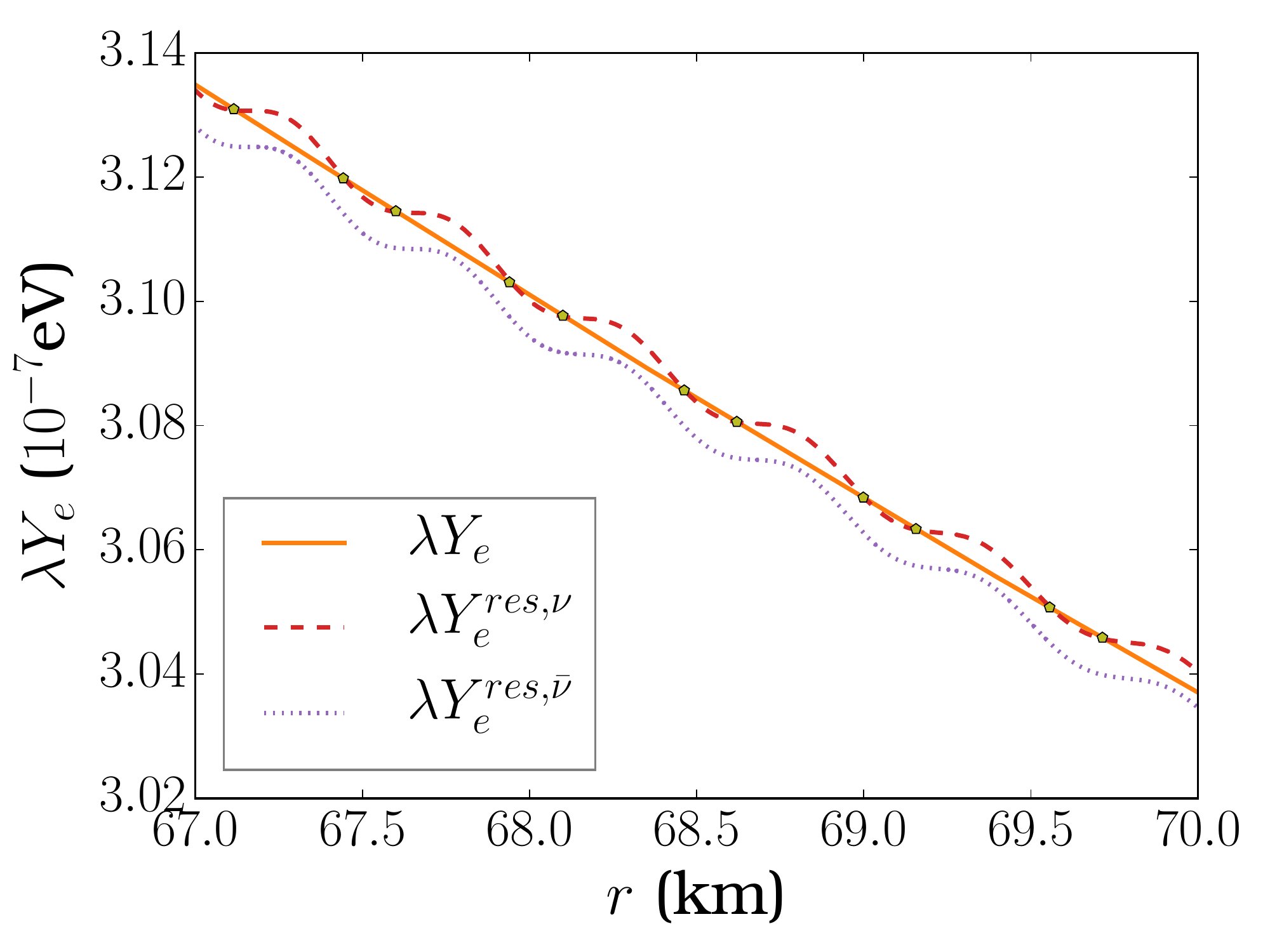}
\caption{Matter profile $\lambda Y_e$ for model A (solid line), and the right-hand side of Eq.\eqn{eq:MNRcondition} for neutrinos (dashed line) and antineutrinos (dotted line). The pentagons show the multiple crossing where NMR resonance condition Eq.\eqn{eq:MNRcondition} is fulfilled.}
\label{fig:yeresMNR}
\end{center}
\end{figure}

\noindent
The MNR phenomenon can maintain over long distances (several hundreds of kilometers) due to a non-linear feedback mechanism that appears because of the self-interaction term.  It involves multiple MSW-like resonances, as  discussed in Refs.\cite{Wu:2015fga,Frensel:2016fge}.
Therefore to maintain the resonant phenomenon,  condition \eqn{e:H12} 
\beq
\lambda Y_e \simeq -\lp h_{\nu \nu}^{ee} - h_{\nu \nu}^{xx} \rp + 2 \omega c_{2\theta},
\label{eq:MNRcondition}
\eeq
has to be encountered several times.
On the left-hand side, the matter profile depends on the distance $r$ and is determined by the model used. On the right-hand side, the self-interaction term depends on the geometrical factors Eqs.\eqn{e:B12}-\eqn{e:Ek}, the conversion probabilities and the neutrino fluxes. Note that for antineutrinos, the vacuum term has  an opposite sign, making the value of the electron density at the resonance location slightly smaller than the one for neutrinos.  In Eq.\eqn{eq:MNRcondition} the difference between the diagonal elements of the self-interaction Hamiltonian can be rewritten as\footnote{Note that, in this section, the dependence on time and energy of the various quantities is not explicitly shown for readability.}   
\begin{align}
 h_{\nu \nu}^{ee} - h_{\nu \nu}^{xx} = \sqrt{2} G_F & \int_0^\infty\!\!\!\! \mathrm dp \left[ \lp 2 \mathcal{P}_{\nu_e\rightarrow \nu_e} -1 \rp \lp G_{\nu_e} j_{\nu_e} -G_{\nu_x} j_{\nu_x} \rp \right. \nonumber \\
 &-\left. \lp 2 \mathcal{P}_{\bar{\nu}_e\rightarrow \bar{\nu}_e} -1 \rp \lp G_{\bar{\nu}_e} j_{\bar{\nu}_e} -G_{\nu_x} j_{\nu_x} \rp \right].
\end{align}
where trace conservation has been used.
Figure \ref{fig:yeresMNR} presents an enlarged region of the matter potential as well as the oscillated self-interaction term $h_{\nu \nu}^{ee} - h_{\nu \nu}^{xx}$ for neutrinos and antineutrinos. This is a typical example of the situations encountered in simulations. One can see that the resonance condition is multiply crossed, a characteristics of a non-linear feedback.

Let us assume that the resonance condition \eqn{eq:MNRcondition} is reached for neutrinos at time $t$, and estimate if it would be possible to encounter it  at time $t+\Delta t$. By assuming that the resonance triggers a small conversion of neutrinos, during the time-lapse $t \rightarrow t+\Delta t$, the electron neutrino survival probability becomes
\beq\label{e:deltap}
 \mathcal{P}_{\nu_e\rightarrow \nu_e} \rightarrow  \mathcal{P}_{\nu_e\rightarrow \nu_e} - \Delta \mathcal{P},
\eeq 
 \noindent
with $\Delta \mathcal{P} >0$\footnote{We assume that the amplitude of the oscillations of the probabilities are small compared to the conversions triggered by the resonance, hence the sign of $\Delta \mathcal{P}$. Note that antineutrinos are not converted since we consider the MNR and not the sMNR.}, while the matter term in \Eqn{eq:MNRcondition} gets
\beq
\lambda Y_e \rightarrow \lambda Y_e + {d{\lp \lambda Y_e \rp} \over dt} \Delta t.
\eeq
On the other hand, the corresponding variation of the self-interaction term in Eq.(\ref{eq:MNRcondition}) includes two contributions 
\begin{align}
\frac{1}{\sqrt{2} G_F} \lp h_{\nu \nu}^{ee} \right.& \left.- h_{\nu \nu}^{x x} \rp \rightarrow \frac{1}{\sqrt{2} G_F} \lp h_{\nu \nu}^{ee} - h_{\nu \nu}^{x x} \rp \nonumber \\
 &- 2   \int_0^\infty\!\!\!\! \mathrm dp \Delta \mathcal{P} \lp G_{\nu_e} j_{\nu_e} -G_{\nu_x} j_{\nu_x} \rp  \nonumber \\
 &+  \Delta t   \int_0^\infty\!\!\!\! \mathrm dp \left[ \lp 2 \mathcal{P}_{\nu_e\rightarrow \nu_e} -1 \rp \lp \dot{G}_{\nu_e} j_{\nu_e} -\dot{G}_{\nu_x} j_{\nu_x} \rp \right. \nonumber \\
 &-\left. \lp 2 \mathcal{P}_{\bar{\nu}_e\rightarrow \bar{\nu}_e} -1 \rp \lp \dot{G}_{\bar{\nu}_e} j_{\bar{\nu}_e} -\dot{G}_{\nu_x} j_{\nu_x} \rp \right].
 \label{eq:selfvar}
\end{align}
The second term, arising from $\Delta \mathcal{P}$, is negative because $G_{\nu_e} j_{\nu_e} > G_{\nu_x} j_{\nu_x}$. As for the third term, in the case of the MNR, $\lp 2 \mathcal{P}_{\nu_e\rightarrow \nu_e} -1 \rp \leq \lp 2 \mathcal{P}_{\bar{\nu}_e\rightarrow \bar{\nu}_e} -1 \rp$ and as it can be observed from Figures \ref{fig:one}-\ref{fig:two}, 
$| \dot{G}_{\bar{\nu}_e} j_{\bar{\nu}_e} -\dot{G}_{\nu_x} j_{\nu_x}| \geq  |  \dot{G}_{\nu_e} j_{\nu_e} -\dot{G}_{\nu_x} j_{\nu_x}|$. Hence, the third term has a positive sign since it is dominated by antineutrinos and the derivatives of the geometric coefficients are always negative. 

To fulfill condition \Eqn{eq:MNRcondition}  again at time $t+\Delta t$,  we need to have a matching between the variation of the self-interaction contribution and the slope of the matter potential. This matching requires the flavor conversions and the decrease of the geometric factors to compensate. For example, for an increasing matter profile, we find that the flavor conversion must have a bigger weight that the decrease of the geometric factor in order to have multiple crossing due to non-linear feedback. 
Let us emphasize that the oscillations of the self-interaction term on the right-hand side of \Eqn{eq:MNRcondition} around the matter potential are possible because these two contributions have opposite signs : hence, they create a yo-yo effect (see Figure \ref{fig:yeresMNR})\footnote{Had they had the same sign, more peculiar conditions would have been needed to get several crossings.}.

Interestingly  we have observed that imposing the non-trace part of $h_{\nu\nu}^{xx}$ to be zero does not prevent the non-linear feedback to happen. Indeed, the same analysis can be repeated and shows that non-linear feedback is still possible. 
 
\subsection{Non-linear feedback in a one-flavor model}
\label{sub:nlfeedvlasenko}
\noindent
Having discussed under which conditions the non-linearity of the equations enables multiple resonances for the MNR, we perform a similar analysis for helicity coherence effects within the model of Ref.\cite{Vlasenko:2014bva}. In fact it is found that a cancellation between the matter and the self-interaction terms occurs over long distances, and a non-linear feedback mechanism produces significant flavor change (depending on the parameters of the model).
Such a model considers only one neutrino flavor, and the associated antineutrino, propagating in a matter background of electrons, (anti)neutrinos and neutrons. Neutrinos traveling along the symmetry axis of a cone interact with those emitted with a fixed angle $\theta=45^{\circ}$.

With these assumptions, the generalized Hamiltonian is 
\beq\label{e:1}
h_\mathcal{G}(t)
= \left(\begin{tabular}{c|c}
$\sqrt{2} G_F n_B Y_e + h_{\nu\nu}^{ee}$  & $\frac{m}{p} h_{\nu\nu}^{\perp, ee}$  \\ 
\hline \\ [-.35cm]
 $\frac{m}{p} \lp h_{\nu\nu}^{\perp, ee} \rp^\dagger$ &  $\sqrt{2} G_F n_B \lp 1- 2Y_e \rp  - h_{\nu\nu}^{ee}$
\end{tabular} \right),
\eeq
where $h_{\nu\nu}^{ee}= 2 \sqrt{2} G_F\lp 1-u \rp \lp n_\nu - n_{\bar{\nu}} \rp $, $u = \cos(\theta)$, with the (anti)neutrino number density  $n_{\nu}$ ($n_{\bar{\nu}}$) and
\begin{align}
n_{\nu} - n_{\bar{\nu}} &= \int\! \mathrm dp\ \phi \lp p  \rp  \rho_{ee} \lp r, p \rp \nonumber \\ 
&= \int\! \mathrm dp\ \phi \lp p \rp   \mathcal{P}_{\nu_e\rightarrow \nu_e} \lp r, p \rp, 
\label{eq:nunubardensity}
\end{align}
$\phi$ being a function that includes the Fermi-Dirac distributions and other numerical factors (which are not relevant here). The off-diagonal term in Eq.\eqn{e:1} is 

\beq
h_{\nu\nu}^{\perp,ee} = 2 \sqrt{2} G_F \sqrt{1-u^2}  \lp n_{\nu} - n_{\bar{\nu}} \rp.
\eeq

With this generalized Hamiltonian, the helicity coherence resonance condition becomes 
\beq
\sqrt{2}G_F n_B \lp 3 Y_e -1 \rp + 2h_{\nu\nu}^{ee} \simeq 0,
\eeq
and is satisfied  if a cancellation between the matter and the self-interaction terms takes place. 
The $Y_e$ value at resonance can be written as 
\beq
Y^{\text{res}}_e  \simeq \frac{1}{3}-\frac{4}{3} \lp 1-u \rp \frac{n_{\nu} - n_{\bar{\nu}}}{n_B}.
\label{eq:rescondvlasenko}
\eeq
In \cite{Vlasenko:2014bva}, it is argued that the neutrino contribution being relatively small, this resonance is located around $Y^0_e = \frac{1}{3}$. In the model, $n_B$ is taken to be a constant while $Y_e$ is increasing according to the profile $Y_e = Y^0_e + \frac{r}{\lambda} \lp 1 + \frac{r^2}{\kappa^2} \rp$, where $\lambda$ and $\kappa$ are two parameters that are allowed to vary. 

Let us perform the same analysis as for the MNR, and suppose that the resonance condition \eqn{eq:rescondvlasenko} has been fulfilled at time $t$, and has triggered a small neutrino conversion  $\mathcal{P}_{\nu_e\rightarrow \nu_e} \rightarrow  \mathcal{P}_{\nu_e\rightarrow \nu_e} - \Delta \mathcal{P}$. Note that, here, $\Delta \mathcal{P}$ is due to a conversion of neutrinos into antineutrinos and vice versa. Then, the lepton number density Eq.\eqn{eq:nunubardensity} decreases  
\beq
n_{\nu} - n_{\bar{\nu}} \rightarrow n_{\nu} - n_{\bar{\nu}} - \Delta n_{\nu-\bar{\nu}},
\eeq
where $\Delta n_{\nu-\bar{\nu}} = \int\! \mathrm dp \ \phi \lp p \rp \Delta \mathcal{P}$.

Therefore the $Y_e$ value at resonance increases according to 
\beq
Y^{\text{res}}_e \rightarrow Y^{\text{res}}_e  + \frac{4}{3} \lp 1-u \rp \frac{\Delta n_{\nu-\bar{\nu}}}{n_B}.
\eeq
Since the chosen $Y_e$ profile increases, it is possible to encounter the resonance more than once. However, as in the case of the MNR, one needs the matching between the slope of $Y_e$ and the conversion $\Delta \mathcal{P}$ of neutrinos into antineutrinos on a short time-scale, that are expected to be small. Therefore, this analysis indicates that, provided that $Y_e$ increases very slowly, the resonance condition can be fulfilled several times\footnote{Note that, since here there is no variations due to geometry, the small oscillations of the survival probabilities are sensible, and lead oscillations of  $Y^{\text{res}}_e$ (see Figure 3 of Ref.\cite{Vlasenko:2013fja}).}.

\subsection{Non-linear feedback and helicity coherence}

Let us now explore the possibility of having a non-linear feedback for the helicity coherence resonance. We study here the resonance condition \eqn{e:H13}, though the discussion can be easily extended to the three other resonance conditions Eqs.\eqn{e:H14}, \eqn{e:H23} and \eqn{e:H24}. The resonance condition \eqn{e:H13} is fulfilled for
\beq
 n_B \lp 3Y_e -1 \rp  \simeq -\frac{2}{\sqrt{2} G_F} h_{\nu \nu}^{ee}.
\label{eq:HC2fcondition}
\eeq
In most cases we have studied, the electron antineutrino contribution   dominates along the trajectories, hence $h_{\nu \nu}^{ee} \leq 0$,  making resonance condition fulfilled for $Y_e \geq \frac{1}{3}$. The self-interaction term  \eqn{e:hexpsa} can be written as
\begin{align} \label{eq:HC2self}
\frac{1}{\sqrt{2} G_F}h_{\nu \nu}^{ee} =   & \int_0^\infty\!\!\!\! \mathrm dp \left[ \lp  \mathcal{P}_{\nu_e\rightarrow \nu_e} + 1 \rp G_{\nu_e} j_{\nu_e} \right. \nonumber \\
& \left. - \lp  \mathcal{P}_{\bar{\nu}_e\rightarrow \bar{\nu}_e} +1 \rp G_{\bar{\nu}_e} j_{\bar{\nu}_e} \right. \nonumber \\
 &+ \left. \lp \mathcal{P}_{\nu_x \rightarrow \nu_e} - \mathcal{P}_{\bar{\nu}_x \rightarrow \bar{\nu}_e} \rp G_{\nu_x} j_{\nu_x}  \right].
\end{align}
We consider the case of Model B where the MNR resonance condition is not met while the helicity coherence one is. 
In this case, $\mathcal{P}_{\nu_x \rightarrow \nu_e}$ and $\mathcal{P}_{\bar{\nu}_x \rightarrow \bar{\nu}_e}$ are frozen and equal to zero, 
while the variations of $\mathcal{P}_{\nu_e\rightarrow \nu_e}$ and $\mathcal{P}_{\bar{\nu}_e\rightarrow \bar{\nu}_e}$ are both equal to $\Delta \mathcal{P}$. 

Let us suppose that the resonant condition \eqn{eq:HC2fcondition} is fulfilled at time $t$ and has triggered conversion of neutrinos into antineutrinos. By using Eq.\eqn{e:deltap}
and a similar relation for antineutrinos,
the self-interaction term varies as \footnote{As for the MNR, we suppose that the small oscillations in the survival probabilities are negligible in comparison with the variations of the geometric coefficients. }
\begin{align}
\frac{1}{\sqrt{2} G_F}h_{\nu \nu}^{ee}  & \rightarrow \frac{1}{\sqrt{2} G_F}h_{\nu \nu}^{ee} 
 - \int_0^\infty\!\!\!\! \mathrm dp \Delta \mathcal{P} \lp G_{\nu_e} j_{\nu_e} - G_{\bar{\nu}_e} j_{\bar{\nu}_e} \rp \nonumber \\
&+\Delta t \int_0^\infty\!\!\!\! \mathrm dp \left[ \lp  \mathcal{P}_{\nu_e\rightarrow \nu_e} + 1 \rp \dot{G}_{\nu_e} j_{\nu_e} \right. \nonumber \\
& \left. - \lp  \mathcal{P}_{\bar{\nu}_e\rightarrow \bar{\nu}_e} +1 \rp \dot{G}_{\bar{\nu}_e} j_{\bar{\nu}_e}  \right].
\end{align}
The contribution due to $\Delta \mathcal{P}$ is positive when antineutrinos dominate the emissions at the neutrinosphere; while the one from the gradient of the geometrical factors are also positive in BNS merger environments. This gives overall a positive sign.
If the matter potential gradient is positive, the matching condition becomes impossible.  On the other hand if the matter gradient is negative, 
peculiar conditions would be necessary to produce oscillations (characteristics of a non-linear feedback mechanism) of $h_{\nu \nu}^{ee}$ around the matter term $\sqrt{2} G_F n_B \lp 3 Y_e -1 \rp$ (similarly to Figure \ref{fig:yeresMNR} for the MNR). 

It can be noticed that even if we had a electron-neutrino dominated environment such as core-collapse supernovae, in which the fulfillment of the resonance condition \eqn{eq:HC2fcondition} would require $Y_e< \frac{1}{3}$, the two contributions to the variation of $h_{\nu \nu}^{ee}$ would still have the same sign\footnote{Unless there are very specific flavor conversions beforehand.}, making it difficult to establish a non-linear feedback mechanism. A different geometry 
with softer geometric factors might make the matching of the two terms  in the helicity resonance condition easier to meet. 

Let us conclude that for such a resonance, a non-linear feedback would enable to increase greatly the adiabaticity. Indeed, using the expression of the adiabaticity parameter $\gamma_m$ introduced in \Eqn{e:adiab}, we find that without a matching of the derivatives of $h_{\mathcal{G},11}$ and $h_{\mathcal{G},33}$, $\gamma_m$ is proportional to $( \frac{m}{q})^2$. For a typical value of $\frac{m}{q} \approx 10^{-7} - 10^{-8}$, we see that this adiabaticity parameter is extremely small. A non-linear feedback would enable the matching of the derivative, and increase $\gamma_m$ up to $\gamma_m = \mathcal{O}( \frac{m}{q})$.

\section{Conclusions}
\noindent
We have explored  the impact of mass contributions  on neutrino flavor evolution in astrophysical environments. These  non-relativistic corrections  appear in extended mean-field descriptions of neutrino propagation. 
We have discussed conditions for the resonances associated with such mass terms and pointed out that, in particular,  they require the matter potential  to be larger than the neutrino self-interaction potentials. 

We have presented the first  study of mass effects in a binary neutron star merger environment. In particular
we have built a two-flavor model based on two-dimensional BNS merger simulations. We have presented numerical results on the neutrino probabilities and adiabaticity during flavor evolution for the following three model cases  where resonance conditions are fulfilled : A) MNR; B) helicity coherence ; C) MNR and helicity coherence. These are representative of the ensemble of results we have been obtaining.
An important result we find is that resonance conditions can be met in  simulations of astrophysical environments such as BNS mergers. However adiabaticity is not sufficient to produce efficient flavor conversion due to helicity coherence. 

It has to be noted that our model is based on the {\it ansatz} that, in the self-interaction Hamiltonian, the flavor evolution of the neutrino modes behave the same as the test neutrino. This approximation gives more weight to the geometrical factor present in the helicity coherence term. Therefore one cannot exclude that the implementation of the full geometrical dependence of the density matrix might introduce some differences with respect to our findings. It is likely, however, that the induced decoherence among the neutrino modes might also not be in favor of  
adiabaticity.

From the present investigation, some general conclusions can be drawn on mass effects. First of all, resonance conditions for helicity coherence can be met  in realistic astrophysical scenarios. On the other hand, the factor $m/q$ suppresses the mass terms values by $10^{-7}-10^{-8}$, if one considers a typical neutrino energy and 0.1 eV as upper limit on the absolute neutrino mass. 
However their role could be magnified by a non-linear feedback mechanism.
We have investigated why multiple crossings, characteristic of such a feedback are absent in our study.
To this aim two cases have been considered where non-linear feedback is operative : the neutrino-matter resonances and the model of Ref.\cite{Vlasenko:2014bva}. 
In fact, in the case of the MNR there is a matching between the derivative of the matter potential and the variations of the self-interaction contribution. Such a matching is possible if the variation arising from flavor contribution and the one arising from the decrease of the geometric factors have the proper weights in order to enable the difference of the self-interaction terms to follow the matter term.  

In the model of Ref.\cite{Vlasenko:2014bva} the signs of the variations on short time-scales still render multiple resonances possibles. 
The adiabaticity being governed by the derivative of the matter term, this matching produces sufficiently adiabatic evolution and a non-linear feedback.
This is in agreement with the results of \cite{Vlasenko:2014bva}, which showed that, for a given value of the mass $m$, $\lambda$ has to be chosen large enough so that the non-linear mechanism can take place. Note that here the non-linear adjustment does not involve geometrical factors.

Our analysis reveals that the MNR and in helicity coherence resonance are essentially of the same nature. Indeed, they both come from the cancellation of a matter term and a self-interaction term. Moreover, the conditions required to trigger a non-linear feedback phenomenon are very similar, though the weighting of the different terms differs. 

For the case of helicity coherence we have argued that the peculiar conditions for multiply crossing the resonance condition are difficult to meet 
because of the strong $r$-dependence of the geometrical factors, of the $\bar{\nu}_e$ over $\nu_e$ dominance in BNS mergers and of the derivative of the matter potentials. 
However our findings also show that, even in a core-collapse environment where $Y_e < 1/3$, multiple resonances would still be difficult to meet under normal conditions. Softer geometric coefficients, met in different environments, could make this matching easier to achieve.

\begin{acknowledgments}
We thank Maik Uwe Frensel for providing us with averaged profiles from the binary neutron star  mergers simulations of Ref.\cite{Perego:2014fma}, for useful discussions and for reading 
the manuscript.
We thank Albino Perego for useful discussions at an early stage of this project 
and Meng-Ru Wu for pointing out an inconsistency in the 
parallel geometrical factor. The authors also acknowledge support
from "Gravitation et physique fondamentale" (GPHYS) of the {\it Observatoire de Paris}.
\end{acknowledgments}

\appendix
\section{Extended evolution equations with mass contributions : Dirac case}
\noindent
In the case neutrinos are Dirac particles, one has to evolve two extended equations including the mass contributions, namely
\begin{equation}\label{e:eqD}
i\, \dot{\!{\rho}}_{D,{\cal G}} \left(t, \vec{q}\, \right)= \left[ h_{D,{\cal G}}\left(t, \vec{q}\, \right), \rho_{D,{\cal G}}\left(t, \vec{q}\, \right) \right],
\end{equation}
and 
\begin{equation}\label{e:eqD2}
i\, \dot{\!\bar{\rho}}_{D,{\cal G}} \left(t, \vec{q}\, \right)= \left[ \bar{h}_{D,{\cal G}}\left(t, \vec{q}\, \right),{\bar{\rho}}_{D,{\cal G}}\left(t, \vec{q}\, \right)\right].
\end{equation}
The explicit expressions of the generalised Hamiltonian in Eq.(\ref{e:matrixform}) is
\beq
\label{e:H}
h_{D,{\cal G}}\left(t, \vec{q}\, \right)
 \!\equiv\!
   \left(\!\!\begin{tabular}{cc}
 $H(t,\vec q\,)$&$\tilde{\Phi}(t,\vec q\,)$\\
 $\tilde{\Phi}^\dagger(t,\vec q\,)$&$\tilde H(t,\vec q\,)$
\end{tabular}\! \! \right),
\eeq
while the generalized density is given by
\beq
\label{eq:dhelicity1}
\rho_{D,{\cal G}}\left(t, \vec{q}\, \right)\equiv
 \left(\begin{tabular}{cc}
 $\rho_{--}(t,\vec q\,)$&$\rho_{-+}(t,\vec q\,)$\\
 $\rho_{+-}(t,\vec q\,)$&$\rho_{++}(t,\vec q\,)$
\end{tabular}\right)\equiv
 \left(\begin{tabular}{cc}
 $\rho(t,\vec q\,)$&$\zeta(t,\vec q\,)$\\
 $\zeta^\dagger(t,\vec q\,)$&$\tilde\rho(t,\vec q\,)$
\end{tabular} \right),
\eeq
where the subscripts in the density matrix $\rho_{--}, \rho_{-+}, \rho_{++}$ indicate the possible
helicity states. In particular, the correlator $\rho_{++}$ refers to 
a sterile state and the $ \rho_{-+}$ couples neutrinos to such 
sterile component. 

For the antineutrino sector, the generalized density is given by
\beq
\bar{\rho}_{D,{\cal G}} \left(t, \vec{q}\, \right) \equiv
   \left(\begin{tabular}{cc}
 $\bar\rho_{--}(t,\vec q\,)$&$\bar\rho_{-+}(t,\vec q\,)$\\
 $\bar\rho_{+-}(t,\vec q\,)$&$\bar\rho_{++}(t,\vec q\,)$
\end{tabular}  \right)\equiv
  \left(\begin{tabular}{cc}
 $\tilde{\bar\rho}(t,\vec q\,)$&$\bar\zeta^\dagger(t,\vec q\,)$\\
 $\bar\zeta(t,\vec q\,)$&$\bar\rho(t,\vec q\,)$
\end{tabular}  \right)\!,
\eeq
with $\bar\rho_{++}$ the usual density matrix for antineutrinos,
$\bar\rho_{--}$ corresponding to a sterile state and $\bar\rho_{-+}$
that couples the sterile with active antineutrino states.
The generalized Hamiltonian in the antineutrino sector
reads 
\beq
\label{e:Hbar}
\bar{ h}_{D,{\cal G}} \left(t, \vec{q}\, \right)
\!\equiv\!
   \left(\!\begin{tabular}{cc}
 $\tilde{\bar H}(t,\vec q\,)$&$\bar\Phi^\dagger(t,\vec q\,)$\\
 $\bar\Phi(t,\vec q\,)$&$\bar H(t,\vec q\,)$
\end{tabular}\! \! \right)\!\!,
\eeq
In the Hamiltonian expressions (\ref{e:H}) and (\ref{e:Hbar}), the off-diagonal terms
couple the neutrinos or antineutrinos  with sterile components, 
as in presence of magnetic fields \cite{Giunti:2014ixa}.

Therefore one gets for the component of $h_{D,{\cal G}}(t)$ Eq.(\ref{e:H}) the following
expressions, by retaining contributions up to order ${\cal O}( m/q)$ from the neutrino mass  in the interaction terms 
\begin{align}
\label{eq:hamcompfirst}
H(t,\vec q\,)&=S(t,q)-\hat q\cdot\vec V(t)-\hat q\cdot \vec V_m(t),\\
 \tilde{\Phi}(t,\vec q\,)&=e^{i\phi_q}\hat\epsilon_q^*\cdot\vec V(t){m\over 2q},\\
 \tilde H(t,\vec q\,)&=h_0(q),
\end{align}
and for $\bar{\rho}_{D,{\cal G}}(t)$ Eq.(\ref{e:Hbar}) 
\begin{align}
\label{eq:hamcompfirstbar}
 \bar H(t,\vec q\,)&=\bar S(t,q)-\hat q\cdot\vec V(t)-\hat q\cdot \vec V_m(t),\\
 \bar\Phi(t,\vec q\,)&=e^{i\phi_q}\hat\epsilon_q\cdot\vec V(t){m\over 2q},\\
\label{eq:hamcomplast}
 \tilde{\bar H}(t,\vec q\,)&=-h_0(q),
\end{align}
The quantities $S(t,q)$, $\bar S(t,q)$ and $\vec V(t)$ are defined in Eqs. \eqn{eq:scalar}, \eqn{eq:scalarbar}, and \eqn{eq:vector} respectively. 
The mass correction to the vector component of the self-interaction Hamiltonian reads
\begin{align}
\label{eq:masscorrecvec}
 \vec V_m(t)&=-\sqrt{2}G_F\!\!\int {\mathrm d^3 {p} \over{(2 \pi)^3}} \Big\{ e^{-i\phi_p}\hat\epsilon_p\,\Omega(t,\vec p\,)\frac{m}{2p}  +  {\rm h.c.}\Big\} \nonumber \\
  & -\sqrt{2}G_F ~{\rm tr}\int {\mathrm d^3 {p} \over{(2 \pi)^3}}  \Big\{ e^{-i\phi_p}\hat \epsilon_p\, \Omega(t,\vec{p}) \frac{m}{2p} +  {\rm h.c.} \Big\},
\end{align}
which gives an extra contribution to the diagonal part of the generalized Hamiltonians.

\section{Geometric factor in the perpendicular self-interaction term}

\begin{figure}[htbp]
\centering
\includegraphics[scale=1]{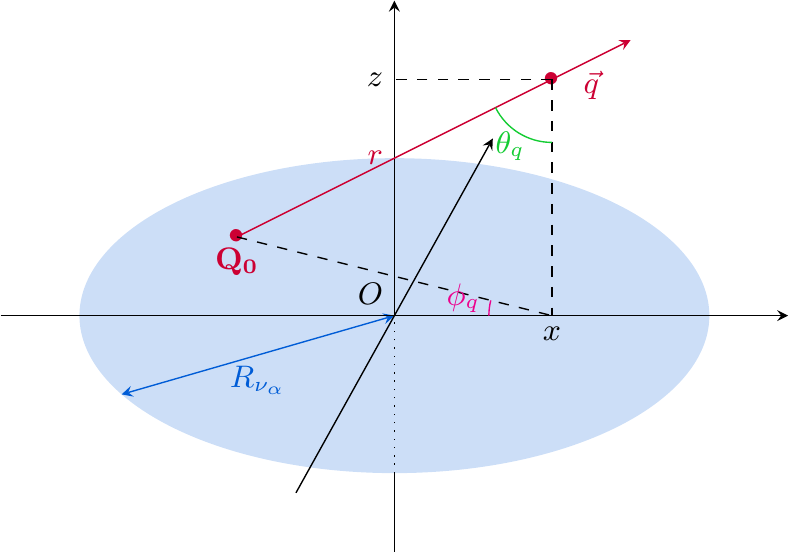}
\caption{Side view of the accretion disk, with the central object located at the center. The radius of the disk is flavor dependent and is denoted $R_{\nu_\alpha}$. A neutrino is emitted near the disk at the point ${\bf Q_0}$, and then leaves it with a momentum $\vec{q}$ located by its spherical coordinates $(q, \theta_q, \phi_q) $. The coordinate $r$ is the distance between the location of the neutrino at time $t$ and its emission point, with the corresponding cartesian coordinates $\lp x, 0, z\rp$.}
\label{fig:def_geometry}
\end{figure}

\noindent
In this section, we compute the geometric factor that is involved in 
\begin{align}
h_{\nu \nu}^\perp (r,q,\ell_q)
&= \sqrt{2} G_F \sum_\alpha \!\! \int_0^{\infty} \!\!\!\!\!\!\! \mathrm dp  \left \{ \int_{\Omega_{\nu_\alpha}} \!\!\!\!\!\!\! \lp \hat{\epsilon}^* \lp \hat{q} \rp \cdot \hat{p} \rp \rho_{\nu_{\underline{\alpha}}} (r,p,\ell_p) \mathrm dn_{\nu_{\underline{\alpha}}} \right. \nonumber \\
&  -\left. \int_{\Omega_{\bar{\nu}_\alpha}} \!\!\!\!\! \!\lp \hat{\epsilon}^* \lp \hat{q} \rp \cdot \hat{p} \rp \bar{\rho}_{\bar{\nu}_{\underline{\alpha}}} (r,p,\ell_p) \mathrm dn _{\bar{\nu}_{\underline{\alpha}}}  \right \},
\label{eq:hselfperp2}
\end{align}
where $\hat{q}$ is the vector of the propagating neutrino, with coordinates $\lp \theta_q, \phi_q \rp$, $\epsilon \lp \hat{q} \rp$ is the unitary vector introduced in Eq.\eqn{eq:lightlike}, $\mathrm dn_{\nu_{\underline{\alpha}}}$ denotes the differential neutrino number density. 
In the spherical coordinates introduced before, $\hat{\epsilon}\lp \hat{q}\rp$ reads 
\beq
\epsilon \lp \hat{q} \rp = \hat{q}_\theta - i\hat{q}_\phi = \begin{pmatrix}
\cos\phi_q \cos\theta_q + i \sin\theta_q \\
 \sin\theta_q  \cos\theta_q -i \cos\phi_q \\
- \sin\theta_q
\end{pmatrix},
\eeq
hence, 
\begin{align}
\epsilon^* \lp \hat{q} \rp \cdot \hat{p} &= \lp\cos\phi_q \cos\theta_q - i \sin\theta_q \rp \stp \cpp  \nonumber \\
&+  \lp \sin\theta_q  \cos\theta_q + i \cos\phi_q  \rp \stp \spp \nonumber \\
&- \sin\theta_q \ctp.
\label{eq:epdotq}
\end{align}

With the approximation given by Eq.(\ref{e:sa}) for the density matrix, the angular integral that needs to be performed is reduced and becomes
\beq\label{e:cperp2}
G^\perp_{\nu_\alpha}\left( r, \ell_q \right) = \int_{\Omega_{\nu_\alpha}} \! (\hat{\epsilon}^* \lp \hat{q})  \cdot \hat{p} \rp \mathrm d\phi_p \mathrm d\cos{\theta_p}.
\eeq
The procedure to perform the angular integral Eq.(\ref{e:cperp2}) is analogous to the case of the geometrical factor appearing in the usual self-interaction Hamiltonian
Eq.(\ref{e:hexp}), as done in Refs.\cite{Malkus:2012ts,Malkus:2015mda,Frensel:2016fge}.

\begin{figure}[htbp]
\centering
\includegraphics[scale=1]{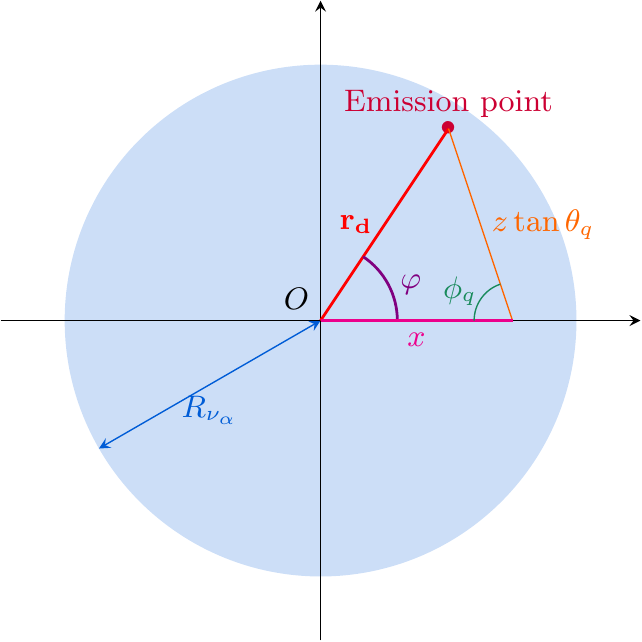}
\caption{Bird's eye view of the accretion disk. The emission point of the neutrino is located by its polar coordinates $\lp r_d, \varphi \rp$.}
\label{fig:birdeye}
\end{figure}

Following Ref.\cite{Dasgupta:2008cu}, we express the variables $\lp \theta_p, \phi_p \rp$ as functions of the polar coordinates of the emission point on the disk, $\lp r_d, \varphi \rp$ (Figure \ref{fig:birdeye}). The following relations 
\beq
\tan{\theta_p} = \frac{1}{z} \sqrt{x^2+r_d^2- 2x r_d \cos{\varphi}},
\eeq

\beq
\cpp = \frac{x- r_d\cos{\varphi}}{\sqrt{x^2+r_d^2-2xr_d\cos{\varphi}}},
\eeq
with $x=x_0+r\stq$ and $z=z_0+r\ctq$, enable to compute the Jacobian of the transformation, leading to 
\begin{align}
\int_{\Omega_{\nu_\alpha}}\!\!\!\!\!\!\mathrm d\phi_p \mathrm d\cos{\theta_p}&  = \nonumber \\
&  \int_0^{R_{\nu_\alpha}} \!\!\!\!\!\mathrm dr_d \int_0^{2\pi}\! \!\!\! \mathrm d\varphi \frac{r_d z}{\lp x^2+z^2+r_d^2-2x r_d \cos{\varphi} \rp^{3/2}}.
\label{eq:jacobian}
\end{align}

Let us define $\Gamma^\perp \lp r_d, r, \ell_q \rp$ such that 
\beq
G_{\nu_\alpha}^\perp \left( r, \ell_q \right) = 
 \int_0^{R_{\nu_\alpha}} \!\!\!\!\!\mathrm dr_d \lp r_d z\rp \Gamma^\perp \lp r_d,r, \ell_q \rp.
 \eeq
 Then, using \eqn{eq:epdotq} and \eqn{eq:jacobian}, the angular integration over $\varphi$ can be performed and leads to :
 \begin{align}\label{e:Gper}
\Gamma^\perp & \lp r_d,r,\ell_q \rp  \nonumber \\
&= \frac{\pi}{\lp ml \rp^{3/2}} \left\lbrace \lp m+l \rp \left[ x \lp \cpq \ctq - i \spq \rp - z \stq \right] \right. \nonumber \\
& \ \ \ \ \left. -4xr_d^2 \lp \cpq \ctq - i \spq \rp \right\rbrace,
 \end{align}
where $m = \lp x+r_d \rp^2+z^2 $ and $l =\lp x-r_d \rp^2+z^2$. Note that the integrals performed here are the same as the ones in the case of the usual self-interaction term Eq.\eqn{e:hexpsa}, but weighted differently. In the case of $\phi_q=0$, $\Gamma^\perp$ is reduced to 
\begin{align}\label{e:Gperp}
\Gamma^\perp & \lp r_d, r, \ell_q \rp \nonumber \\
& = \frac{\pi}{\lp ml \rp^{3/2}} \left[ \lp m+l \rp \lp x \ctq - z \stq \rp -4xr_d^2 \ctq \right].
\end{align}

As for the geometrical factor along the neutrino direction of motion, one has
\beq\label{e:B12}
G_{\nu_\alpha} \left( r, \ell_q \right) = 
 \int_0^{R_{\nu_\alpha}} \!\!\!\!\!\mathrm dr_d \lp r_d z\rp \Gamma \lp r_d,r, \ell_q \rp.
 \eeq
where we define $\Gamma  \lp r_d, r, \ell_q \rp$ similarly to $\Gamma^\perp$.
It also involves angular integrals, with different weights as in the case of the perpendicular term 
\begin{align} \label{eq:Gamma_factor}
\Gamma \lp r_{d},r, \ell_q \rp & =  \int_0^{2\pi} \!\!\!\!\! \mathrm d\varphi  \left[ 1- \stq \stp  \lp \cpq \cpp \right. \right. \nonumber \\
& \left. \left.  + \spq \spp  \rp - \ctq \ctp  \right] \nonumber \\
&  \times \frac{1}{\lp x^2+z^2+r_d^2-2xr_d\cos{\varphi}\rp^{3/2}},
\end{align}
where the angles $\theta_p$ and $\phi_p$ must be expressed as functions of $\lp r_q, \varphi\rp$. 
The explicit $\varphi$-integration in \Eqn{eq:Gamma_factor} yields, for $\phi_q=0$ 
\begin{align}\label{e:geomg}
\Gamma &\lp r_{d},r, \ell_q \rp   = \frac{4}{l\sqrt{m}} E\lp \sqrt{\frac{m-l}{m}} \rp \nonumber \\
&- \frac{\pi}{\lp ml \rp^{3/2}} \left[ \lp m+l \rp \lp z\ctq + x \stq\rp- 4x r_d^2 \stq\right],
\end{align}
where the relation $m - l = 4 x r_{\mathrm{d}}$ has been used, with $m$ and $l$ defined previously, and $E(k)$ denotes Legendre's complete elliptic integral of the second kind 
\begin{equation}\label{e:Ek}
E(k) \equiv \int_{0}^{\pi/2}\!\!\!\! \mathrm{d} \theta \, \sqrt{1 - k^{2} \sin^{2}{\theta}},
\end{equation}
where we have extended the usual definition domain from $k \in \left[ 0,1 \right]$ to $k \in \left[ 0,1 \right] \cap i \mathds{R} $. Note that $-\Gamma / 2$ with the replacement $\phi_q \mapsto \pi - \phi_q$ corresponds to the geometric factor $C$ given in \cite{Malkus:2012ts}. 
Note  also the different convention used to denote the elliptic integral.

\section{Adiabaticity}
\label{an:adiab}
\noindent
We remind that in the $\rm SU (2)$ isospin formalism the equations of motion are replaced by precession equations where an effective magnetic field is built from the
Hamiltonian. In the two flavor case it is given by  
\beq\label{e:B}
\vec{B} = \left(\begin{tabular}{c}
$ 2~ {\rm Re}(H_{ex})$\\$ -2 {\rm ~Im}(H_{ex})$\\ $H_{ee} - H_{xx}$ \end{tabular}\right),
\eeq
and the effective isospins are constructed from the density matrices 
\beq\label{e:S}
\vec{P} = {1 \over 2}\left(\begin{tabular}{c}
$ 2 ~{\rm Re}(\rho_{ex})$\\$ -2 {\rm ~Im}(\rho_{ex})$\\ $\rho_{ee} - \rho_{xx}$ \end{tabular}\right),
\eeq
The third component of the isospin vectors gives information on the flavor content, while the $x$- and $y$-components of the isospins contain the mixings.  

In our analysis of mass effects, 
we consider that at the helicity coherence resonance, flavor conversions are frozen, which is well 
justified when MNR and the helicity coherence resonances are separated or the MNR is ineffective. Hence, the system is effectively 
reduced to $2\times2$ corresponding to electron neutrinos and electron antineutrinos. We can then define the effective magnetic field as
\beq\label{e:Bm}
\vec{B}_m = \left(\begin{tabular}{c}
$ 2~ {\rm Re}(h_{\mathcal{G},13})$\\$ -2 {\rm ~Im}(h_{\mathcal{G},13})$\\ $h_{\mathcal{G},11} - h_{\mathcal{G},33}$ \end{tabular}\right),
\eeq
and the effective isospin 
\beq
\vec{P}_m = {1 \over 2}\left(\begin{tabular}{c}
$ 2 ~{\rm Re}(\rho^{-+}_{ee})$\\$ -2 {\rm ~Im}(\rho^{-+}_{ee})$\\ $\rho^{--}_{ee} - \rho^{++}_{ee}$ \end{tabular}\right).
\eeq

Within this formalism, the MSW resonance condition corresponds to the third component of the magnetic field being zero while the evolution is adiabatic if the precession of the isospins is fast compared to the rate of change of the magnetic field. 
In this case, the isospins manage to keep approximately  alignment with the effective magnetic fields, so that the cosine of the angle between the total isospin and the magnetic fields remains similar before and after the resonance. 
Another way of quantifying adiabaticity of the evolution is through the gamma factor
\beq\label{e:adiab}
\gamma =\frac{| \vec{B}| ^3}{\left| \frac{\mathrm d \vec{B}} {\mathrm dt} \times  \vec{B} \right|}.
\eeq
where $\vec{B}$ stands for $\vec{B}$ \eqn{e:B} or $\vec{B}_m$ \eqn{e:Bm} in our notations.
If $\gamma \gg 1$ evolution is adiabatic.

\end{document}